\documentclass[aps,pre,showkeys,
longbibliography,
reprint,                
twocolumn,              
nofootinbib,
floatfix]{revtex4-1}

\usepackage[T1]{fontenc}
\usepackage[utf8]{inputenc}
\usepackage{amsmath}
\usepackage{amsfonts}
\usepackage{graphicx}
\usepackage{tikz}
\usepackage[labelformat=simple,justification=justified]{subcaption}

\usepackage{algorithm}
\usepackage{algpseudocode}
\usepackage{dcolumn}
\usepackage{csquotes}

\usepackage{listings}
\definecolor{codegreen}{rgb}{0,0.6,0}
\definecolor{codegray}{rgb}{0.5,0.5,0.5}
\definecolor{codepurple}{rgb}{0.58,0,0.82}
\definecolor{backcolour}{rgb}{0.95,0.95,0.92}
\lstdefinestyle{mystyle}{
    backgroundcolor=\color{backcolour},   
    commentstyle=\color{codegreen},
    keywordstyle=\color{magenta},
    numberstyle=\tiny\color{codegray},
    stringstyle=\color{codepurple},
    basicstyle=\ttfamily\footnotesize,
    breakatwhitespace=false,         
    breaklines=true,                 
    captionpos=t,                    
    keepspaces=true,                 
    numbers=left,                    
    numbersep=5pt,                  
    showspaces=false,                
    showstringspaces=false,
    showtabs=false,                  
    tabsize=2
}
\usepackage[colorlinks=true,hyperfootnotes=true,breaklinks=true]{hyperref}
\usepackage[capitalise]{cleveref}

\begin{document}

\title{Random site percolation thresholds on square lattice for complex neighborhoods containing sites up to the sixth coordination zone}
\author{Krzysztof Malarz}
\thanks{\href{https://orcid.org/0000-0001-9980-0363}{0000-0001-9980-0363}}
\email{malarz@agh.edu.pl}
\affiliation{AGH University, Faculty of Physics and Applied Computer Science, al.~Mickiewicza~30, 30-059 Krak\'ow, Poland}

\begin{abstract}
The site percolation problem is one of the core topics in statistical physics.
Evaluation of the percolation threshold, which separates two phases (sometimes described as conducting and insulating), is useful for a range of problems from core condensed matter to interdisciplinary application of statistical physics in epidemiology or other transportation or connectivity problems. 
In this paper with Newman--Ziff fast Monte Carlo algorithm and finite-size scaling theory the random site percolation thresholds $p_c$ for a square lattice with complex neighborhoods containing sites from the sixth coordination zone are computed. 
Complex neighborhoods are those that contain sites from various coordination zones (which are not necessarily compact).
We also present the source codes of the appropriate procedures (written in C) to be replaced in original Newman--Ziff code.
Similar to results previously found for the honeycomb lattice, the percolation thresholds for complex neighborhoods on a square lattice follow the power law $p_c(\zeta)\propto\zeta^{-\gamma_2}$ with $\gamma_2=0.5454(60)$, where $\zeta=\sum_i z_i r_i$ is the weighted distance of sites in complex neighborhoods ($r_i$ and $z_i$ are the distance from the central site and the number of sites in the coordination zone $i$, respectively).
\end{abstract}

\date{October 23, 2023}


\maketitle

\section{Introduction}

Percolation \cite{Broadbent1957,Hammersley1957} is one of the core problems in statistical physics with many interdisciplinary applications ranging from
materials science \cite{ISI:000514848600043},
through studies of polymer composites \cite{ISI:000528948100007},
forest fires \cite{Kaczanowska2002}, 
agriculture \cite{ISI:000518460000003},
oil and gas exploration \cite{ISI:000524118200031}, 
diseases propagation \cite{2101.00550}, 
transportation networks \cite{ISI:000528691800009},
quantifying urban areas \cite{ISI:000523958600016},
to Bitcoins transfer \cite{Bartolucci2020} (see References~\onlinecite{Li_2021,Saberi2015} for reviews).
The percolating system undergoes a (purely geometrical) phase transition (in terms of the conductivity or transportation properties of the system) from the phase corresponding to an insulator (for low connectivity $p<p_c$) to a conductor (for high connectivity $p>p_c$). 
The critical connectivity of the system $p_c$ (called the percolation threshold) separates these two phases and depends on the dimension of the system $d$, the topology of the lattice, the number $z$ of sites in the assumed neighborhood, the type of percolation (that is, the site or bond dilution), etc. \cite{bookDS,Wierman2014}.

Percolation thresholds were initially estimated for nearest-neighbor interactions \cite{Dean_1963,Dean_Bird_1967,PhysRevE.60.275} but later also complex neighborhoods (termed also extended for compact neighborhoods) were studied for various lattices embedded in:
\begin{itemize}
\item $d=2$ (for a square \cite{Dalton_1964,Domb1966,Gouker1983,Galam2005a,Galam2005b,Majewski2007,PhysRevE.103.022126}, 
a triangular \cite{Dalton_1964,Domb1966,Iribarne1999,2006.15621,2102.10066}, 
a honeycomb \cite{Dalton_1964,2204.12593} 
and other Archimedean \cite{Lebrecht_2021,PhysRevE.105.024105} lattices);
\item $d=3$ (for a simple cubic \cite{Kurzawski2012,Malarz2015,PhysRevE.103.022126} lattice);
\item $d=4$ (for a simple hypercubic \cite{1803.09504,Zhao_2022} lattice)
\end{itemize}
dimensions.

\begin{figure*}[htbp]
\begin{subfigure}[b]{0.14\textwidth}
\caption{\label{fig:sq-1nn}}
\begin{tikzpicture}[scale=0.36]
\draw[step=1,color=gray] (-3.5,-3.5) grid (3.5,3.5);
\draw[very thick,red] (0, 0) circle (.4);
\draw[ultra thick,orange] (0,0) circle (1);
\filldraw[black] (0,-1) circle (.4);
\filldraw[black] (0,+1) circle (.4);
\filldraw[black] (-1,0) circle (.4);
\filldraw[black] (+1,0) circle (.4);
\end{tikzpicture}
\end{subfigure}
\hfill
\begin{subfigure}[b]{0.14\textwidth}
\caption{\label{fig:sq-2nn}}
\begin{tikzpicture}[scale=0.36]
\draw[step=1,color=gray] (-3.5,-3.5) grid (3.5,3.5);
\draw[very thick,red] (0, 0) circle (.4);
\draw[ultra thick,orange] (0,0) circle (1.414213562);
\filldraw[black] (+1,-1) circle (.4);
\filldraw[black] (+1,+1) circle (.4);
\filldraw[black] (-1,-1) circle (.4);
\filldraw[black] (-1,+1) circle (.4);
\end{tikzpicture}
\end{subfigure}
\hfill
\begin{subfigure}[b]{0.14\textwidth}
\caption{\label{fig:sq-3nn}}
\begin{tikzpicture}[scale=0.36]
\draw[step=1,color=gray] (-3.5,-3.5) grid (3.5,3.5);
\draw[very thick,red] (0, 0) circle (.4);
\draw[ultra thick,orange] (0,0) circle (2);
\filldraw[black] ( 0,-2) circle (.4);
\filldraw[black] ( 0,+2) circle (.4);
\filldraw[black] (-2, 0) circle (.4);
\filldraw[black] (+2, 0) circle (.4);
\end{tikzpicture}
\end{subfigure}
\hfill
\begin{subfigure}[b]{0.14\textwidth}
\caption{\label{fig:sq-4nn}}
\begin{tikzpicture}[scale=0.36]
\draw[step=1,color=gray] (-3.5,-3.5) grid (3.5,3.5);
\draw[very thick,red] (0, 0) circle (.4);
\draw[ultra thick,orange] (0,0) circle (2.2360679);
\filldraw[black] (-2,-1) circle (.4);
\filldraw[black] (-2,+1) circle (.4);
\filldraw[black] (-1,-2) circle (.4);
\filldraw[black] (-1,+2) circle (.4);
\filldraw[black] (+1,-2) circle (.4);
\filldraw[black] (+1,+2) circle (.4);
\filldraw[black] (+2,-1) circle (.4);
\filldraw[black] (+2,+1) circle (.4);
\end{tikzpicture}
\end{subfigure}
\hfill
\begin{subfigure}[b]{0.14\textwidth}
\caption{\label{fig:sq-5nn}}
\begin{tikzpicture}[scale=0.36]
\draw[step=1,color=gray] (-3.5,-3.5) grid (3.5,3.5);
\draw[very thick,red] (0, 0) circle (.4);
\draw[ultra thick,orange] (0,0) circle (2.8284271);
\filldraw[black] (-2,-2) circle (.4);
\filldraw[black] (-2,+2) circle (.4);
\filldraw[black] (+2,-2) circle (.4);
\filldraw[black] (+2,+2) circle (.4);
\end{tikzpicture}
\end{subfigure}
\hfill
\begin{subfigure}[b]{0.14\textwidth}
\caption{\label{fig:sq-6nn}}
\begin{tikzpicture}[scale=0.36]
\draw[step=1,color=gray] (-3.5,-3.5) grid (3.5,3.5);
\draw[very thick,red] (0, 0) circle (.4);
\draw[ultra thick,orange] (0,0) circle (3);
\filldraw[black] (0,-3) circle (.4);
\filldraw[black] (0,+3) circle (.4);
\filldraw[black] (-3,0) circle (.4);
\filldraw[black] (+3,0) circle (.4);
\end{tikzpicture}
\end{subfigure}
\caption{\label{fig:sq-basic-neighbourhoods}Shapes of basic neighborhoods on square lattice. 
\subref{fig:sq-1nn} \textsc{sq}-1, $r^2=1$,
\subref{fig:sq-2nn} \textsc{sq}-2, $r^2=2$,
\subref{fig:sq-3nn} \textsc{sq}-3, $r^2=4$,
\subref{fig:sq-4nn} \textsc{sq}-4, $r^2=5$,
\subref{fig:sq-5nn} \textsc{sq}-5, $r^2=8$,
\subref{fig:sq-6nn} \textsc{sq}-6, $r^2=9$.
The number $r$ is the radius of (orange) circle indicating equidistant sites marked by solid (black) circles to the central one marked with open (red) circle}
\end{figure*}

Simultaneously with the estimation of percolation thresholds for various lattices, some effort went into searching for an analytical formula allowing for the prediction of the percolation threshold position based on lattice characteristics.
For example, \citeauthor{PhysRevE.105.024105} \cite{PhysRevE.105.024105} estimated the site and bond percolation thresholds for 11 Archimedean lattices with complex and compact (extended) neighborhoods containing sites up to the tenth coordination zone.
For the site percolation problem, the critical site occupation probability $p_c$ follows asymptotically
\begin{equation}
\label{eq:pc-discs}
p_c(z) = a/z
\end{equation}
with the total number $z$ of sites in the neighborhood and $a\approx 4.51235$.
This dependence should be reached exactly for the percolation of compact neighborhoods with a large number $z$ of sites that make up the neighborhood (for example, for discs).
To take into account finite-$z$ effect an additional term $b$ in the denominator of \Cref{eq:pc-discs}
\begin{equation}
\label{eq:pc-finite-z}
p_c(z) = c/(z+b)
\end{equation}
has been included \cite{PhysRevE.103.022127}.
For the two-dimensional lattices $b=3$ \cite{PhysRevE.105.024105}.
The third universal scaling studied in by \citeauthor{PhysRevE.105.024105} \cite{PhysRevE.105.024105} was
\begin{equation}
\label{eq:pc-exp}
p_c(z;d) = 1-\exp(d/z)
\end{equation}
proposed by \citeauthor{Koza_2014} \cite{Koza_2014,Koza_2016}.

Much earlier \citeauthor{PhysRevE.53.2177} \cite{PhysRevE.53.2177,PhysRevE.55.1230} proposed a universal formula for site percolation problem 
\begin{equation}
\label{eq:pc-GM}
p_c(z;d)= \dfrac{p_0}{[(d-1)(z-1)]^a}.
\end{equation}
They recognized two classes of systems (two sets of $(p_0,a)$ parameters) \cite{PhysRevE.53.2177}.
Their paper \cite{PhysRevE.53.2177} was immediately criticized by \citeauthor{ISI:A1997WD54600080} \cite{ISI:A1997WD54600080} who showed 
`an example of two networks, where $d$ and $z$ are equal, but the percolation thresholds differ'.

For complex neighborhoods, the situation is even more complex, since for a given lattice topology (and thus fixed $d$) there are many neighborhoods with exactly the same total number $z$ of sites in the neighborhood but different percolation thresholds $p_c$ (see: Table 1 and Figure 4 in Reference \onlinecite{Majewski2007} for the square lattice; Table 1 in Reference \onlinecite{2006.15621} and Table 1 and Figure 3(a) in Reference \onlinecite{2102.10066} for the triangular lattice; and Table 1 and Figure 4(a) in Reference \onlinecite{2204.12593} for the honeycomb lattice).

To solve the above-mentioned problems of $p_c(z)$ degeneration the index 
\begin{equation} \label{eq:xi} 
\xi=\sum_i z_i r_i^2/i
\end{equation}
was proposed by \citeauthor{2102.10066} \cite{2102.10066}.
The $z_i$ and $r_i$ are the number of sites and their distance from the central site in the neighborhood in the $i$-th coordination zone. 
The index $\xi$ allowed for a successful distinguishing between neighborhoods and cancel $p_c(z)$ degeneration for the triangular lattice with complex neighborhoods containing sites up to the fifth coordination zone.
The dependence of the percolation threshold 
\begin{equation} \label{eq:pc_vs_xi} 
p_c(\xi)\propto\xi^{-\gamma_1}
\end{equation}
was well fitted with the power law with $\gamma_1(\textsc{tr})\approx 0.710(19)$.
Unfortunately, this dependence does not hold for the honeycomb lattice (see Figure 4(b) in Reference \onlinecite{2204.12593}). 
Thus, another index
\begin{equation} \label{eq:zeta} 
\zeta=\sum_i z_i r_i
\end{equation}
was introduced by \citeauthor{2204.12593}, to simultaneously resolve the problem of $p_c(z)$ degeneration and to distinguish among various complex neighborhoods for the honeycomb lattice  \cite{2204.12593}. 
For honeycomb lattice and complex neighborhoods up to the fifth coordination zone 
\begin{equation} \label{eq:pc_vs_zeta} 
p_c(\zeta)\propto\zeta^{-\gamma_2}
\end{equation}
with $\gamma_2(\textsc{hc})\approx 0.4981(90)$ \cite{2204.12593}.

In this paper, using the fast Monte Carlo Newman--Ziff algorithm \cite{NewmanZiff2001}, we calculate the critical occupation probabilities $p_c$ (percolation thresholds) for random site percolation in a square lattice and neighborhoods combined with basic neighborhoods presented in \Cref{fig:sq-basic-neighbourhoods}.
The basic neighborhoods contain sites from the first coordination zone (\textsc{sq-1}, \Cref{fig:sq-1nn}) up to the sixth coordination zone (\textsc{sq-6}, \Cref{fig:sq-6nn}).
These complex neighborhoods are presented in \Cref{fig:sq-neighbourhoods} in \Cref{app:shapes}.
Calculations of percolation thresholds are based on the finite-size scaling hypothesis \cite{Finite-Size_Scaling_Theory_1990,bookDS,Guide_to_Monte_Carlo_Simulations_2009}.

The second aim of this paper is to check if \Cref{eq:pc_vs_xi,eq:pc_vs_zeta} holds for a square lattice with complex neighborhoods and, if so, which of them performs better.

The rest of the paper is organized as follows. The details of the calculations are presented in the following \Cref{sec:computations}.
The results of the calculations are given in \Cref{sec:results}.
The article is summarized and concluded in \Cref{sec:conclusions}.
The \Cref{app:shapes} contains graphical presentation of neighborhood shapes.
In ``Supplementary materials'' we present: 
\begin{itemize}
    \item a set of {\tt boundaries()} functions (written in C, 
    Listings 1 to 6)
    to be replaced in the Newman--Ziff program published in Reference \onlinecite{NewmanZiff2001} to obtain the single realization of $\mathcal S_{\max}(n;L)$ for the neighborhoods presented in \Cref{fig:sq-1nn,fig:sq-2nn,fig:sq-3nn,fig:sq-4nn,fig:sq-5nn,fig:sq-6nn};
    \item and the dependencies of $\mathcal{P}_{\max}\cdot L^{\beta/\nu}$ on the probability of occupation $p$ for neighborhoods ranging from \textsc{sq-6} to \textsc{sq-1,2,3,4,5,6} for various linear system sizes $L=128$ to 4096.
\end{itemize}

\section{\label{sec:computations}Computations}

Our calculations of the percolation thresholds $p_c$ are based on finite-size analyses of the probability $\mathcal{P}_\text{max}$ that the randomly selected site belongs to the largest cluster of occupied sites.
According to the finite-size hypothesis \cite{Finite-Size_Scaling_Theory_1990,bookDS,Guide_to_Monte_Carlo_Simulations_2009}, in the vicinity of a phase transition (marked by a critical point $x_c$), many quantity $\mathcal A$ characterizing the system obeys a scaling relation
\begin{equation}
\mathcal A(x;L)=L^{-\varepsilon_1} \mathcal F \left( (x-x_c)L^{\varepsilon_2} \right),
\end{equation}
where $x$ measures the level of system disorder (temperature for the Ising or Potts model, site/bond occupation probability for percolation problem), $L$ is the linear size of the system, $\mathcal F$ is a scaling function (usually analytically unknown) and $\varepsilon_1$ and $\varepsilon_2$ are scaling exponents.
In other words, there exists a function $\mathcal F$, that for properly assumed values of $x_c$, $\varepsilon_1$ and $\varepsilon_2$ the dependencies of $\mathcal A(x)$ collapse into a single curve independently of the (finite) system size $L$.
This also provides an elegant way to predict the value of the critical parameter $x_c$ as 
\begin{equation}
\mathcal A(x;L) L^{\varepsilon_1} = \mathcal F \left( (x-x_c)L^{\varepsilon_2} \right),
\end{equation}
which for $x=x_c$ yields
\begin{equation}
\mathcal A(x_c;L) L^{\varepsilon_1} = \mathcal F \left( 0 \right).
\end{equation}
In other words, we expect the curves $L^{\varepsilon_1} \mathcal A(x)$ plotted for various sizes of linear systems $L$ to intercept each other at $x=x_c$.

For our purposes, we assume that $\mathcal A\equiv \mathcal P_\text{max}$ (the probability that a randomly selected site belongs to the largest cluster) and $x\equiv p$ (the probability of occupation of the sites).
For the problem of site percolation, the critical values of the exponents $\varepsilon_1$ and $\varepsilon_2$ are known exactly \cite[p.~54]{bookDS} as $\varepsilon_1=\frac{5}{36}/\frac{4}{3}=\frac{5}{48}$ and $\varepsilon_2=1/\frac{4}{3}=\frac{3}{4}$.

To compute the probability of belonging to the largest cluster
\begin{equation}
\label{eq:Pmax}
\mathcal{P}_{\max}(p;L)=\mathcal S_\text{max}(p;L)/N
\end{equation}
we first need to calculate the sizes of the largest cluster $S_\text{max}$ and $N=L^2$ is the number of all sites available in the system.

To that end, we use three concepts presented in Reference \onlinecite{NewmanZiff2001}.
\begin{itemize}
\item The first is the fast system construction scheme (known as the Newman--Ziff algorithm).
The efficiency of this approach is based on the recursive construction of the system with $n$ occupied sites with the addition of only one occupied site to the system containing $(n-1)$ already occupied sites.
\item The second concept is the way of transforming the $\bar{\mathcal A}(n;N)$ dependence on the integer number of occupied sites $n$ into the dependence $\mathcal A(p;N)$ on the probability of the site occupation $p$
\begin{equation}
\label{eq:An-to-Ap}
\mathcal A(p;N)=\sum_{n=0}^N \bar{\mathcal A}(n;N) \mathcal B(n;N,p),
\end{equation}
where
\begin{equation}
\label{eq:binom}
\mathcal B(n;N,p)=\binom{N}{n}p^n(1-p)^{N-n}
\end{equation}
are values of the binomial (Bernoulli) probability distribution.
\item The third concept is the efficient construction of the binomial coefficients \eqref{eq:binom}.
\end{itemize}

The applied scheme defined in \Cref{eq:An-to-Ap} together with the construction of the binomial distribution coefficients is presented in \Cref{alg:Qn2Qp}.

\begin{algorithm}[H]
\caption{\label{alg:Qn2Qp}Conversion $\bar{\mathcal A}(n)$ to $\mathcal A(p)$ \cite{NewmanZiff2001}}
\begin{algorithmic}[1]
\Require $p_1$, $p_2$, $\Delta p$, $N$, $\bar{\mathcal A}(n)$ \Comment{$n\in\{0,1,\cdots,N-1,N\}$} 
\Ensure $\mathcal A(p)$ \Comment{for $p$ from $p_1$ to $p_2$ every $\Delta p$}
\State $p\gets p_1$
\While{$p\le p_2$} 
    \State $n_\text{max}=pN$ \Comment{store $\mathcal B(N,n,p)$ to $\hat B$}
    \State $B(n_\text{max})=pN$
    \For{$n=n_\text{max}+1,N$}
        \State $B(n)=B(n-1)\cdot\dfrac{(N-n+1)p}{n(1-p)}$
    \EndFor
    \For{$n=n_\text{max}-1,0,-1$}
        \State $B(n)=B(n+1)\cdot\dfrac{(n+1)(1-p)}{p(N-n)}$
    \EndFor
    \State $c\gets\sum\hat B$
    \State $\hat B\gets\hat B/c$
    \State $\mathcal A(p)=0$ 
    \ForAll{$n$}
        \State $\mathcal A(p)\gets \mathcal A(p)+B(n)\bar{\mathcal A}(n)$
    \EndFor
    \State \textbf{return} $p,\mathcal A(p)$ 
    \State $p\gets p+\Delta p$
\EndWhile
\end{algorithmic}
\end{algorithm}

\section{\label{sec:results}Results}

\begin{figure}
\begin{subfigure}[b]{0.49\textwidth}
\caption{\label{fig:Smaxvsn-SQ-123456-example}}
\includegraphics[width=0.89\textwidth]{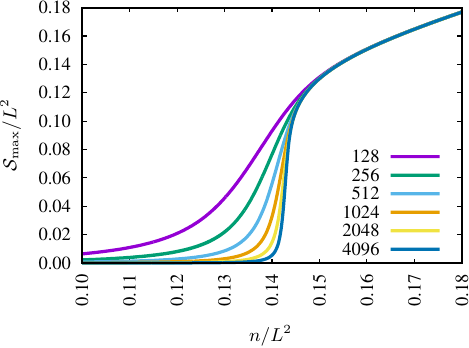}
\end{subfigure}
\begin{subfigure}[b]{0.49\textwidth}
\caption{\label{fig:PmaxLbetanuvsp-SQ-123456-example-1}}
\includegraphics[width=0.89\textwidth]{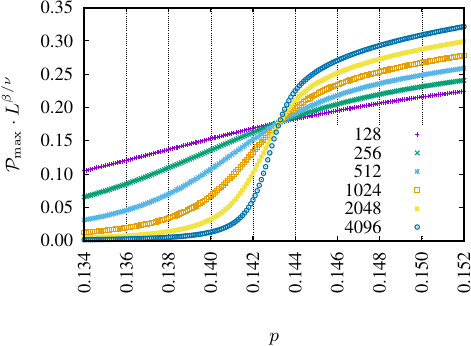}
\end{subfigure}
\begin{subfigure}[b]{0.49\textwidth}
\caption{\label{fig:PmaxLbetanuvsp-SQ-123456-example-2}}
\includegraphics[width=0.89\textwidth]{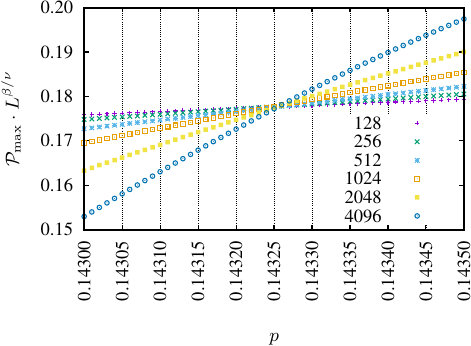}
\end{subfigure}
\caption{\label{fig:example-results}An example of results obtained for \textsc{sq-1,2,3,4,5,6} neighborhood and various linear system sizes $L=128$ to 4096.
\subref{fig:Smaxvsn-SQ-123456-example} The size of the largest cluster $\mathcal{S}_\text{max}$ vs. the number of occupied sites $n$. Both quantities are normalized to the system size $L^2$.
\subref{fig:PmaxLbetanuvsp-SQ-123456-example-1} Dependence of $L^{\varepsilon_1} \mathcal P_\text{max}(p)$ for $p$ from 0.134 to 0.152 every $\Delta p=10^{-4}$.
\subref{fig:PmaxLbetanuvsp-SQ-123456-example-2} Close-up on \Cref{fig:PmaxLbetanuvsp-SQ-123456-example-1} in the vicinity of the percolation threshold $p_c$ for $p$ from 0.1430 to 0.1435 every $\Delta p=10^{-5}$}
\end{figure}

In \Cref{fig:example-results} examples of the results (for neighborhood \textsc{sq-1,2,3,4,5,6} and various sizes of linear systems $L=128$, 256, 512, 1024, 2048, and 4096) of the computations obtained with the procedure described in \Cref{sec:computations} are presented.
\Cref{fig:Smaxvsn-SQ-123456-example} shows dependencies of the largest cluster size $\mathcal{S}_\text{max}$ (normalized to the system size $L^2$) vs. the number of occupied sites $n$ (also normalized to the system size $L^2$).
With increasing system linear size $L$ the dependence $\mathcal S_\text{max}(n)$ becomes steeper and steeper.
\Cref{fig:PmaxLbetanuvsp-SQ-123456-example-1} shows $L^{\varepsilon_1} \mathcal P_\text{max}(p)$ for $p$ ranging from 0.134 to 0.152 estimated for every $\Delta p=10^{-4}$.
\Cref{fig:PmaxLbetanuvsp-SQ-123456-example-2} shows close-up of \Cref{fig:PmaxLbetanuvsp-SQ-123456-example-1} in the vicinity of the percolation threshold (for $p$ from 0.1430 to 0.1435 for every $\Delta p=10^{-5}$).
The finite-size effects in $L^{\varepsilon_1} \mathcal P_\text{max}(p)$ vanish for $p=p_c$, resulting in a common point of $L^{\varepsilon_1} \mathcal P_\text{max}(p)$ plotted for various linear sizes $L$ of the systems.
The analogous dependencies for all other complex neighborhoods---presented in \Cref{fig:sq-neighbourhoods} in \Cref{app:shapes}---containing sites from the sixth coordination zone are shown in Figure 5. 
The results are averaged over the realizations of the $R=10^5$ system.
The obtained $p_c$ are gathered in \Cref{tab:sq-pc}.

\begin{table}[tbp]
\caption{\label{tab:sq-pc}Percolation thresholds $p_c$ for a square lattice with complex neighborhoods (and their characteristics $z$, $\zeta$, $\xi$) containing sites from the sixth coordination zone}
\begin{ruledtabular}
\begin{tabular}{ldddd}
	lattice & \text{$z$} & \text{$\zeta$} & \text{$\xi$} & \text{$p_c$}\\
\hline 
\hline 
\textsc{sq-1,2,3,4,5,6} & 28	&	58.8591	&	35.7333	&	0.14326\footnote{0.142 \cite{Iribarne1999}, 0.143255 \cite{PhysRevE.103.022126}}\\
\hline 
\textsc{sq-2,3,4,5,6} &	24	&	54.8591	&	31.7333	&	0.14575	\\
\textsc{sq-1,3,4,5,6} &	24	&	53.2023	&	31.7333	&	0.14801	\\
\textsc{sq-1,2,4,5,6} &	24	&	50.8591	&	30.4	&	0.15223	\\
\textsc{sq-1,2,3,5,6} &	20	&	40.9706	&	25.7333	&	0.16661	\\
\textsc{sq-1,2,3,4,6} &	24	&	47.5454	&	29.3333	&	0.16134	\\
\hline 
\textsc{sq-3,4,5,6} &	20	&	49.2023	&	27.7333	&	0.15221	\\
\textsc{sq-2,4,5,6} &	20	&	46.8591	&	26.4	&	0.15844	\\
\textsc{sq-2,3,5,6} &	16	&	36.9706	&	21.7333	&	0.17601	\\
\textsc{sq-2,3,4,6} &	20	&	43.5454	&	25.3333	&	0.16529	\\
\textsc{sq-1,4,5,6} &	20	&	45.2023	&	26.4	&	0.15815	\\
\textsc{sq-1,3,5,6} &	16	&	35.3137	&	21.7333	&	0.18007	\\
\textsc{sq-1,3,4,6} &	20	&	41.8885	&	25.3333	&	0.16675	\\
\textsc{sq-1,2,5,6} &	16	&	32.9706	&	20.4	&	0.18216	\\
\textsc{sq-1,2,4,6} &	20	&	39.5454	&	24	    &	0.17409	\\
\textsc{sq-1,2,3,6} &	16	&	29.6569	&	19.3333	&	0.20134	\\
\hline 
\textsc{sq-4,5,6} &	16	&	41.2023	&	22.4	&	0.16819	\\
\textsc{sq-3,5,6} &	12	&	31.3137	&	17.7333	&	0.19867	\\
\textsc{sq-3,4,6} &	16	&	37.8885	&	21.3333	&	0.17288	\\
\textsc{sq-2,5,6} &	12	&	28.9706	&	16.4	&	0.20036	\\
\textsc{sq-2,4,6} &	16	&	35.5454	&	20	    &	0.18454	\\
\textsc{sq-2,3,6} &	12	&	25.6569	&	15.3333	&	0.21503	\\
\textsc{sq-1,5,6} &	12	&	27.3137	&	16.4	&	0.19936	\\
\textsc{sq-1,4,6} &	16	&	33.8885	&	20	    &	0.18143	\\
\textsc{sq-1,3,6} &	12	&	24	    &	15.3333	&	0.22577	\\
\textsc{sq-1,2,6} &	12	&	21.6569	&	14	    &	0.23076	\\
\hline 
\textsc{sq-5,6} &	8	&	23.3137	&	12.4	&	0.24422	\\
\textsc{sq-4,6} &	12	&	29.8885	&	16	    &	0.19799	\\
\textsc{sq-3,6} &	8	&	20	    &	11.3333	&	0.25673	\\
\textsc{sq-2,6} &	8	&	17.6569	&	10	    &	0.26600	\\
\textsc{sq-1,6} &	8	&	16	    &	10	    &	0.27309	\\
\hline 
\textsc{sq-6}\footnote{equivalent to \textsc{sq-1}} &	4	&	12	&	6	&	0.59274\footnote{0.592746 \cite[p. 17]{bookDS}, 0.59274621(13) \cite{PhysRevLett.85.4104}, 0.59274(5) \cite{PhysRevE.103.012115} for \textsc{sq-1}}\\
\end{tabular}
\end{ruledtabular}
\end{table}

\begin{figure}[tbph]
\begin{subfigure}[b]{0.49\textwidth}
\caption{\label{fig:pc_vs_z}}
\includegraphics[width=0.81\columnwidth]{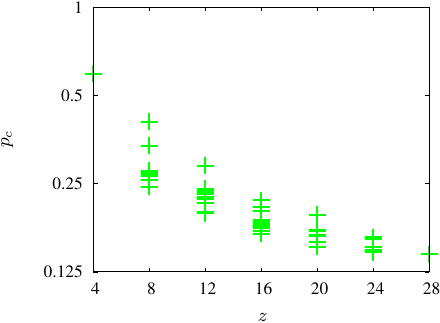}
\end{subfigure}
\begin{subfigure}[b]{0.49\textwidth}
\caption{\label{fig:pc_vs_xi}}
\includegraphics[width=0.81\columnwidth]{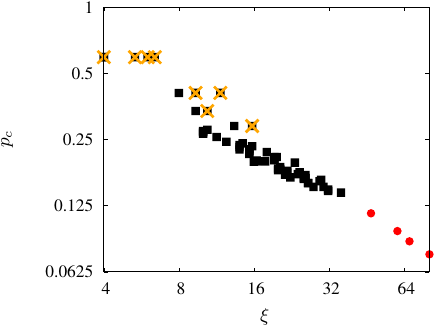}
\end{subfigure}
\begin{subfigure}[b]{0.49\textwidth}
\caption{\label{fig:pc_vs_zeta}}
\includegraphics[width=0.81\columnwidth]{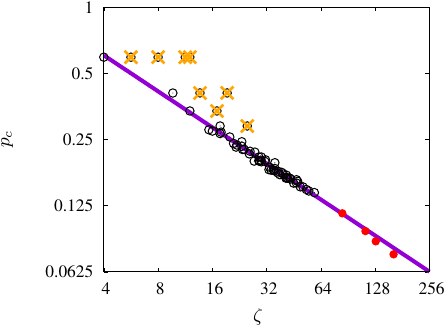}
\end{subfigure}
\caption{\label{fig:pc-vs-z-xi-zeta}\subref{fig:pc_vs_z} Degeneracy of $p_c(z)$ for a square lattice with complex neighborhoods ranging from \textsc{sq}-1 up to \textsc{sq}-1,2,3,4,5,6. 
\subref{fig:pc_vs_xi} Dependence $p_c(\xi)$ for various neighborhoods containing sites up to the sixth coordination zone.
Orange crosses show equivalent neighborhoods (\textsc{sq}-1$\equiv$\textsc{sq}-2$\equiv$\textsc{sq}-3$\equiv$\textsc{sq}-5$\equiv$\textsc{sq}-6,
\textsc{sq}-1,3$\equiv$\textsc{sq}-2,5,
\textsc{sq}-1,2$\equiv$\textsc{sq}-2,3$\equiv$\textsc{sq}-3,5
and \textsc{sq}-1,2,3$\equiv$\textsc{sq}-2,3,5).
\subref{fig:pc_vs_zeta} Dependence $p_c(\zeta)$. Inflated neighborhoods corresponding to higher indexes (i.e., \textsc{sq}-2, \textsc{sq}-3, \textsc{sq}-5, \textsc{sq}-6, \textsc{sq}-2,5, \textsc{sq}-2,3, \textsc{sq}-3,5 and \textsc{sq}-2,3,5) are excluded from the fitting procedure.
The solid (violet) line indicates \Cref{eq:pc_vs_zeta} with $\gamma_2(\textsc{sq})=0.5454(60)$}
\end{figure}

In \Cref{fig:pc-vs-z-xi-zeta} the dependencies of $p_c$ on the total number $z=\sum_i z_i$ of sites in the complex neighborhoods containing the sites of the $i$-th coordination zone and the indexes $\xi$ and $\zeta$ are presented.
\Cref{fig:pc_vs_z} shows the dependence of the percolation threshold $p_c$ on $z$.
The percolation thresholds for neighborhoods containing sites up to the fifth coordination zone are taken from References \onlinecite{Galam2005a,Majewski2007} and those for neighborhoods containing sites from the sixth coordination zone presented here in \Cref{tab:sq-pc}.
It is clear that $z$ cannot differentiate between the various shapes of the neighborhoods or the percolation thresholds associated with them. 

In \Cref{fig:pc_vs_xi} the dependence \eqref{eq:pc_vs_xi} on the percolation threshold $p_c$ for complex neighborhoods on the square lattice on the index $\xi$ is presented.
Similarly to the earlier observation for the honeycomb lattice \cite{2204.12593}, some deviations from the straight line in \Cref{eq:pc_vs_xi} are observed.
The full circles mark percolation thresholds for compact neighbourhoods \textsc{sq}-1,2,$\cdots$,6,7, \textsc{sq}-1,2,$\cdots$,7,8, \textsc{sq}-1,2,$\cdots$,8,9 and \textsc{sq}-1,2,$\cdots$,9,10) taken from References \onlinecite{PhysRevE.103.022126} and \onlinecite{Iribarne1999}.

\Cref{fig:pc_vs_zeta} shows the dependence \eqref{eq:pc_vs_zeta} of $p_c$ for complex neighborhoods on the square lattice on the index $\zeta$.
The percolation thresholds for neighborhoods containing sites up to the fifth coordination zone are taken from References \onlinecite{Galam2005a,Majewski2007}, those for neighborhoods containing sites from the sixth coordination zone presented here in \Cref{tab:sq-pc} and 
those for compact neighborhoods containing sites from the seventh to the tenth coordination zones are taken from References \onlinecite{PhysRevE.103.022126} and \onlinecite{Iribarne1999}.

\section{\label{sec:discussion}Discussion}

The percolation thresholds $p_c$ obtained in simulations range from 0.59275 (for \textsc{sq-6}) to 0.14325 (for \textsc{sq-1,2,3,4,5,6}). 
The latter agrees in five significant digits with its earlier estimate $p_c(\textsc{sq-1,2,3,4,5,6})=0.143255$ \cite{PhysRevE.103.022126}.
The \textsc{sq-6} neighborhood is topologically equivalent to \textsc{sq-1} (but for a three-times larger lattice constant), resulting in identical percolation thresholds $p_c(\textsc{sq-6})=p_c(\textsc{sq-1})$.
For the \textsc{sq-6} neighborhood we deal with several simultaneous independent percolation problems on several identically shaped lattices.
The latter reduces effective system size, but our results show, that this effect is perfectly compensated by effective increase of number of samples.

Similarly to the honeycomb lattice \cite{2204.12593}, the power law \eqref{eq:pc_vs_zeta} also holds for a square lattice with complex neighborhoods with $\gamma_2(\textsc{sq})=0.5454(60)$ given by the least-squares method.
Inflated neighborhoods \textsc{sq-2}, \textsc{sq-3}, \textsc{sq-5}, \textsc{sq-6},
\textsc{sq-2,3}, \textsc{sq-3,5},
\textsc{sq-2,5} 
and \textsc{sq-2,3,5}
(corresponding to \textsc{sq-1}, \textsc{sq-1}, \textsc{sq-1}, \textsc{sq-1}, 
\textsc{sq-1,2}, \textsc{sq-1,2},
\textsc{sq-1,3} 
and \textsc{sq-1,2,3}, respectively) were excluded from the fitting procedure.

Among the neighborhoods that contain sites up to the sixth coordination zone, there are seven pairs of various neighborhoods with exactly the same $\zeta$ index, namely 
$\zeta(\textsc{sq-2,4,5,6}) =\zeta(\textsc{sq-1,2,3,4,5}) \approx 46.86$, 
$\zeta(\textsc{sq-4,5,6})   =\zeta(\textsc{sq-1,3,4,5})   \approx 41.20$, 
$\zeta(\textsc{sq-2,5,6})   =\zeta(\textsc{sq-1,2,3,5})   \approx 28.97$, 
$\zeta(\textsc{sq-2,4,6})   =\zeta(\textsc{sq-1,2,3,4})   \approx 35.55$, 
$\zeta(\textsc{sq-5,6})     =\zeta(\textsc{sq-1,3,5})     \approx 23.31$, 
$\zeta(\textsc{sq-2,6})     =\zeta(\textsc{sq-1,2,3})     \approx 17.66$ and 
$\zeta(\textsc{sq-6}) 	    =\zeta(\textsc{sq-1,3})       = 12$.

The differentiate power of a scalar index $\zeta$ is still better than the differentiate power of an index $\xi$ and both are much better than the differentiate power of the total number of sites in the neighborhood $z$.

\section{\label{sec:conclusions}Conclusion}

In this paper with Newman and Ziff effective Monte Carlo algorithm we calculated percolation thresholds for 32 complex neighbourhoods (containing sites from the sixth coordination zone) on a square lattice.

As scalar indexes $\xi$ \eqref{eq:xi} and $\zeta$ \eqref{eq:zeta} allow simultaneously to (more or less effective) distinguish between various neighbourhoods and accordingly fitting $p_c$ to the inverse power-law on $\xi$ \eqref{eq:pc_vs_xi} or $\zeta$ \eqref{eq:pc_vs_zeta} (with various efficiency depending on the underlying lattice shape) searching for another index remains an open task.
The index may involve various powers of $z_i$, $r_i$ and $i$, where $i$ stands for the number of coordination zone from which sites constituting neighbourhood come from and $z_i$ and $r_i$ are the number and the distance of sites in this coordination zone to the central site in the neighbourhood, respectively.

Also calculating $p_c$ for neighbourhoods containing sites up to the sixth coordination zone on triangular and honeycomb lattices seems to be desired. 
Simultaneous calculation of $p_c$ for honeycomb and triangular lattices should allow for identifying inflated and equivalent neighbourhoods but among these two underlying regular lattices. 
Preliminary inspection of the $p_c(\zeta)$ dependence---but for neighbourhoods containing sites up to the fifth coordination zone \cite{2102.10066}---reveals $\gamma_2$ close to 1/2.
This does not make factor $\xi$ totally useless, as for the bond-percolation problem very clear dependnce \eqref{eq:pc_vs_xi} with $\gamma_1\approx 1$ was recently observed \cite{Xun_2022}.

Finally, the further studies may focus on the fractal nature of the giant component at $p=p_c$ \cite{fractalfract7030231}.
The largest percolating cluster on square lattice at $p=p_c$ has fractal properties for \textsc{sq-1} neighbourhoods with fractal dimension close to 1.9 \cite[p.~9]{bookDS}.
Does this picture survive changing neighbourhoods to complex one, also on other lattice topologies? 
And if yes, is the fractal dimension the same as for the nearest-neighbours interactions?

The results obtained in this paper may be helpful in further searching for the universal formula, in the spirit of \Cref{eq:pc-GM} \cite{Galam_Mauger_1994}, for percolation threshold $p_c$, also for complex neighborhoods, but independently of the underlying two-dimensional lattice shape.
Also further studies on the topic presented here may result in finding universal formula for $p_c$ not only for two-dimensional lattices but in higher dimensions (including nonphysical dimensions, like on four- \cite{1803.09504,Zhao_2022} and five-dimensional simple hyper-cubic \cite{PhysRevE.98.022120,2308.15719} lattices).

\begin{acknowledgments}
I am grateful to Ma{\l}gorzata J. Krawczyk for a fruitful discussion on boundary conditions for neighborhoods containing sites up-to the sixth coordination zone on the square lattice.
I gratefully acknowledge Poland’s high-performance computing infrastructure PLGrid (HPC Centers: ACK Cyfronet AGH) for providing computer facilities and support within computational grant no.~PLG/2023/016295.
\end{acknowledgments}

\appendix

\section{\label{app:shapes}Neighborhoods shapes}

In \Cref{fig:sq-neighbourhoods} the shapes of all complex neighborhoods containing sites from the sixth coordination zone are presented.

\bibliography{percolation,km,basics}

\begin{thebibliography}{51}%
\makeatletter
\providecommand \@ifxundefined [1]{%
 \@ifx{#1\undefined}
}%
\providecommand \@ifnum [1]{%
 \ifnum #1\expandafter \@firstoftwo
 \else \expandafter \@secondoftwo
 \fi
}%
\providecommand \@ifx [1]{%
 \ifx #1\expandafter \@firstoftwo
 \else \expandafter \@secondoftwo
 \fi
}%
\providecommand \natexlab [1]{#1}%
\providecommand \enquote  [1]{``#1''}%
\providecommand \bibnamefont  [1]{#1}%
\providecommand \bibfnamefont [1]{#1}%
\providecommand \citenamefont [1]{#1}%
\providecommand \href@noop [0]{\@secondoftwo}%
\providecommand \href [0]{\begingroup \@sanitize@url \@href}%
\providecommand \@href[1]{\@@startlink{#1}\@@href}%
\providecommand \@@href[1]{\endgroup#1\@@endlink}%
\providecommand \@sanitize@url [0]{\catcode `\\12\catcode `\$12\catcode
  `\&12\catcode `\#12\catcode `\^12\catcode `\_12\catcode `\%12\relax}%
\providecommand \@@startlink[1]{}%
\providecommand \@@endlink[0]{}%
\providecommand \url  [0]{\begingroup\@sanitize@url \@url }%
\providecommand \@url [1]{\endgroup\@href {#1}{\urlprefix }}%
\providecommand \urlprefix  [0]{URL }%
\providecommand \Eprint [0]{\href }%
\providecommand \doibase [0]{http://dx.doi.org/}%
\providecommand \selectlanguage [0]{\@gobble}%
\providecommand \bibinfo  [0]{\@secondoftwo}%
\providecommand \bibfield  [0]{\@secondoftwo}%
\providecommand \translation [1]{[#1]}%
\providecommand \BibitemOpen [0]{}%
\providecommand \bibitemStop [0]{}%
\providecommand \bibitemNoStop [0]{.\EOS\space}%
\providecommand \EOS [0]{\spacefactor3000\relax}%
\providecommand \BibitemShut  [1]{\csname bibitem#1\endcsname}%
\let\auto@bib@innerbib\@empty
\bibitem [{\citenamefont {Broadbent}\ and\ \citenamefont
  {Hammersley}(1957)}]{Broadbent1957}%
  \BibitemOpen
  \bibfield  {author} {\bibinfo {author} {\bibfnamefont {S.~R.}\ \bibnamefont
  {Broadbent}}\ and\ \bibinfo {author} {\bibfnamefont {J.~M.}\ \bibnamefont
  {Hammersley}},\ }\bibfield  {title} {\enquote {\bibinfo {title} {Percolation
  processes: I. {C}rystals and mazes},}\ }\href {\doibase
  10.1017/S0305004100032680} {\bibfield  {journal} {\bibinfo  {journal}
  {Mathematical Proceedings of the Cambridge Philosophical Society}\ }\textbf
  {\bibinfo {volume} {53}},\ \bibinfo {pages} {629--641} (\bibinfo {year}
  {1957})}\BibitemShut {NoStop}%
\bibitem [{\citenamefont {Hammersley}(1957)}]{Hammersley1957}%
  \BibitemOpen
  \bibfield  {author} {\bibinfo {author} {\bibfnamefont {J.~M.}\ \bibnamefont
  {Hammersley}},\ }\bibfield  {title} {\enquote {\bibinfo {title} {Percolation
  processes: {II}. {T}he connective constant},}\ }\href {\doibase
  10.1017/S0305004100032692} {\bibfield  {journal} {\bibinfo  {journal}
  {Mathematical Proceedings of the Cambridge Philosophical Society}\ }\textbf
  {\bibinfo {volume} {53}},\ \bibinfo {pages} {642--645} (\bibinfo {year}
  {1957})}\BibitemShut {NoStop}%
\bibitem [{\citenamefont {Cheng}\ \emph {et~al.}(2020)\citenamefont {Cheng},
  \citenamefont {Yan}, \citenamefont {Yang}, \citenamefont {Zou}, \citenamefont
  {Yang},\ and\ \citenamefont {Liang}}]{ISI:000514848600043}%
  \BibitemOpen
  \bibfield  {author} {\bibinfo {author} {\bibfnamefont {L.}~\bibnamefont
  {Cheng}}, \bibinfo {author} {\bibfnamefont {P.}~\bibnamefont {Yan}}, \bibinfo
  {author} {\bibfnamefont {X.}~\bibnamefont {Yang}}, \bibinfo {author}
  {\bibfnamefont {H.}~\bibnamefont {Zou}}, \bibinfo {author} {\bibfnamefont
  {H.}~\bibnamefont {Yang}}, \ and\ \bibinfo {author} {\bibfnamefont
  {H.}~\bibnamefont {Liang}},\ }\bibfield  {title} {\enquote {\bibinfo {title}
  {High conductivity, percolation behavior and dielectric relaxation of hybrid
  {ZIF}-8/{CNT} composites},}\ }\href {\doibase 10.1016/j.jallcom.2020.154132}
  {\bibfield  {journal} {\bibinfo  {journal} {Journal of Alloys and Compounds}\
  }\textbf {\bibinfo {volume} {825}},\ \bibinfo {pages} {154132} (\bibinfo
  {year} {2020})}\BibitemShut {NoStop}%
\bibitem [{\citenamefont {Zhang}\ \emph {et~al.}(2020)\citenamefont {Zhang},
  \citenamefont {Zhang}, \citenamefont {Guo}, \citenamefont {Guo},\ and\
  \citenamefont {Yu}}]{ISI:000528948100007}%
  \BibitemOpen
  \bibfield  {author} {\bibinfo {author} {\bibfnamefont {Q.}~\bibnamefont
  {Zhang}}, \bibinfo {author} {\bibfnamefont {B.-Y.}\ \bibnamefont {Zhang}},
  \bibinfo {author} {\bibfnamefont {B.-H.}\ \bibnamefont {Guo}}, \bibinfo
  {author} {\bibfnamefont {Z.-X.}\ \bibnamefont {Guo}}, \ and\ \bibinfo
  {author} {\bibfnamefont {J.}~\bibnamefont {Yu}},\ }\bibfield  {title}
  {\enquote {\bibinfo {title} {High-temperature polymer conductors with
  self-assembled conductive pathways},}\ }\href {\doibase
  10.1016/j.compositesb.2020.107989} {\bibfield  {journal} {\bibinfo  {journal}
  {Composites Part B---Engineering}\ }\textbf {\bibinfo {volume} {192}},\
  \bibinfo {pages} {{107989}} (\bibinfo {year} {2020})}\BibitemShut {NoStop}%
\bibitem [{\citenamefont {Malarz}\ \emph {et~al.}(2002)\citenamefont {Malarz},
  \citenamefont {Kaczanowska},\ and\ \citenamefont
  {Ku{\l}akowski}}]{Kaczanowska2002}%
  \BibitemOpen
  \bibfield  {author} {\bibinfo {author} {\bibfnamefont {K.}~\bibnamefont
  {Malarz}}, \bibinfo {author} {\bibfnamefont {S.}~\bibnamefont {Kaczanowska}},
  \ and\ \bibinfo {author} {\bibfnamefont {K.}~\bibnamefont {Ku{\l}akowski}},\
  }\bibfield  {title} {\enquote {\bibinfo {title} {Are forest fires
  predictable?}}\ }\href {\doibase 10.1142/S0129183102003760} {\bibfield
  {journal} {\bibinfo  {journal} {International Journal of Modern Physics C}\
  }\textbf {\bibinfo {volume} {13}},\ \bibinfo {pages} {1017--1031} (\bibinfo
  {year} {2002})}\BibitemShut {NoStop}%
\bibitem [{\citenamefont {Ram\'irez}\ \emph {et~al.}(2020)\citenamefont
  {Ram\'irez}, \citenamefont {Pajares}, \citenamefont {Mart\'inez},
  \citenamefont {Rodr\'iguez~Fern\'andez}, \citenamefont {Molina-Gayosso},
  \citenamefont {Lozada-Lechuga},\ and\ \citenamefont
  {Fern\'andez~T\'ellez}}]{ISI:000518460000003}%
  \BibitemOpen
  \bibfield  {author} {\bibinfo {author} {\bibfnamefont {J.~E.}\ \bibnamefont
  {Ram\'irez}}, \bibinfo {author} {\bibfnamefont {C.}~\bibnamefont {Pajares}},
  \bibinfo {author} {\bibfnamefont {M.~I.}\ \bibnamefont {Mart\'inez}},
  \bibinfo {author} {\bibfnamefont {R.}~\bibnamefont
  {Rodr\'iguez~Fern\'andez}}, \bibinfo {author} {\bibfnamefont
  {E.}~\bibnamefont {Molina-Gayosso}}, \bibinfo {author} {\bibfnamefont
  {J.}~\bibnamefont {Lozada-Lechuga}}, \ and\ \bibinfo {author} {\bibfnamefont
  {A.}~\bibnamefont {Fern\'andez~T\'ellez}},\ }\bibfield  {title} {\enquote
  {\bibinfo {title} {Site-bond percolation solution to preventing the
  propagation of \emph{{P}hytophthora zoospores} on plantations},}\ }\href
  {\doibase 10.1103/PhysRevE.101.032301} {\bibfield  {journal} {\bibinfo
  {journal} {Physical Review E}\ }\textbf {\bibinfo {volume} {101}},\ \bibinfo
  {pages} {032301} (\bibinfo {year} {2020})}\BibitemShut {NoStop}%
\bibitem [{\citenamefont {Ghanbarian}\ \emph {et~al.}(2020)\citenamefont
  {Ghanbarian}, \citenamefont {Liang},\ and\ \citenamefont
  {Liu}}]{ISI:000524118200031}%
  \BibitemOpen
  \bibfield  {author} {\bibinfo {author} {\bibfnamefont {B.}~\bibnamefont
  {Ghanbarian}}, \bibinfo {author} {\bibfnamefont {F.}~\bibnamefont {Liang}}, \
  and\ \bibinfo {author} {\bibfnamefont {H.-H.}\ \bibnamefont {Liu}},\
  }\bibfield  {title} {\enquote {\bibinfo {title} {Modeling gas relative
  permeability in shales and tight porous rocks},}\ }\href {\doibase
  10.1016/j.fuel.2020.117686} {\bibfield  {journal} {\bibinfo  {journal}
  {Fuel}\ }\textbf {\bibinfo {volume} {272}},\ \bibinfo {pages} {117686}
  (\bibinfo {year} {2020})}\BibitemShut {NoStop}%
\bibitem [{\citenamefont {Ziff}(2021)}]{2101.00550}%
  \BibitemOpen
  \bibfield  {author} {\bibinfo {author} {\bibfnamefont {R.~M.}\ \bibnamefont
  {Ziff}},\ }\bibfield  {title} {\enquote {\bibinfo {title} {Percolation and
  the pandemic},}\ }\href {\doibase 10.1016/j.physa.2020.125723} {\bibfield
  {journal} {\bibinfo  {journal} {Physica A: Statistical Mechanics and its
  Applications}\ }\textbf {\bibinfo {volume} {568}},\ \bibinfo {pages} {125723}
  (\bibinfo {year} {2021})}\BibitemShut {NoStop}%
\bibitem [{\citenamefont {Dong}\ \emph {et~al.}(2020)\citenamefont {Dong},
  \citenamefont {Mostafizi}, \citenamefont {Wang}, \citenamefont {Gao},\ and\
  \citenamefont {Li}}]{ISI:000528691800009}%
  \BibitemOpen
  \bibfield  {author} {\bibinfo {author} {\bibfnamefont {S.}~\bibnamefont
  {Dong}}, \bibinfo {author} {\bibfnamefont {A.}~\bibnamefont {Mostafizi}},
  \bibinfo {author} {\bibfnamefont {H.}~\bibnamefont {Wang}}, \bibinfo {author}
  {\bibfnamefont {J.}~\bibnamefont {Gao}}, \ and\ \bibinfo {author}
  {\bibfnamefont {X.}~\bibnamefont {Li}},\ }\bibfield  {title} {\enquote
  {\bibinfo {title} {Measuring the topological robustness of transportation
  networks to disaster-induced failures: {A} percolation approach},}\ }\href
  {\doibase 10.1061/(ASCE)IS.1943-555X.0000533} {\bibfield  {journal} {\bibinfo
   {journal} {Journal of Infrastructure Systems}\ }\textbf {\bibinfo {volume}
  {26}},\ \bibinfo {pages} {04020009} (\bibinfo {year} {2020})}\BibitemShut
  {NoStop}%
\bibitem [{\citenamefont {Cao}\ \emph {et~al.}(2020)\citenamefont {Cao},
  \citenamefont {Dong}, \citenamefont {Wu},\ and\ \citenamefont
  {Liu}}]{ISI:000523958600016}%
  \BibitemOpen
  \bibfield  {author} {\bibinfo {author} {\bibfnamefont {W.}~\bibnamefont
  {Cao}}, \bibinfo {author} {\bibfnamefont {L.}~\bibnamefont {Dong}}, \bibinfo
  {author} {\bibfnamefont {L.}~\bibnamefont {Wu}}, \ and\ \bibinfo {author}
  {\bibfnamefont {Y.}~\bibnamefont {Liu}},\ }\bibfield  {title} {\enquote
  {\bibinfo {title} {Quantifying urban areas with multi-source data based on
  percolation theory},}\ }\href {\doibase 10.1016/j.rse.2020.111730} {\bibfield
   {journal} {\bibinfo  {journal} {Remote Sensing of Environment}\ }\textbf
  {\bibinfo {volume} {241}},\ \bibinfo {pages} {111730} (\bibinfo {year}
  {2020})}\BibitemShut {NoStop}%
\bibitem [{\citenamefont {Bartolucci}\ \emph {et~al.}(2020)\citenamefont
  {Bartolucci}, \citenamefont {Caccioli},\ and\ \citenamefont
  {Vivo}}]{Bartolucci2020}%
  \BibitemOpen
  \bibfield  {author} {\bibinfo {author} {\bibfnamefont {S.}~\bibnamefont
  {Bartolucci}}, \bibinfo {author} {\bibfnamefont {F.}~\bibnamefont
  {Caccioli}}, \ and\ \bibinfo {author} {\bibfnamefont {P.}~\bibnamefont
  {Vivo}},\ }\bibfield  {title} {\enquote {\bibinfo {title} {A percolation
  model for the emergence of the {B}itcoin {L}ightning {N}etwork},}\ }\href
  {\doibase 10.1038/s41598-020-61137-5} {\bibfield  {journal} {\bibinfo
  {journal} {Scientific Reports}\ }\textbf {\bibinfo {volume} {10}},\ \bibinfo
  {pages} {4488} (\bibinfo {year} {2020})}\BibitemShut {NoStop}%
\bibitem [{\citenamefont {Li}\ \emph {et~al.}(2021)\citenamefont {Li},
  \citenamefont {Liu}, \citenamefont {L\"u}, \citenamefont {Hu}, \citenamefont
  {Xu},\ and\ \citenamefont {Zhang}}]{Li_2021}%
  \BibitemOpen
  \bibfield  {author} {\bibinfo {author} {\bibfnamefont {M.}~\bibnamefont
  {Li}}, \bibinfo {author} {\bibfnamefont {R.-R.}\ \bibnamefont {Liu}},
  \bibinfo {author} {\bibfnamefont {L.}~\bibnamefont {L\"u}}, \bibinfo {author}
  {\bibfnamefont {M.-B.}\ \bibnamefont {Hu}}, \bibinfo {author} {\bibfnamefont
  {S.}~\bibnamefont {Xu}}, \ and\ \bibinfo {author} {\bibfnamefont {Y.-C.}\
  \bibnamefont {Zhang}},\ }\bibfield  {title} {\enquote {\bibinfo {title}
  {Percolation on complex networks: {T}heory and application},}\ }\href
  {\doibase 10.1016/j.physrep.2020.12.003} {\bibfield  {journal} {\bibinfo
  {journal} {Physics Reports}\ }\textbf {\bibinfo {volume} {907}},\ \bibinfo
  {pages} {1--68} (\bibinfo {year} {2021})}\BibitemShut {NoStop}%
\bibitem [{\citenamefont {Saberi}(2015)}]{Saberi2015}%
  \BibitemOpen
  \bibfield  {author} {\bibinfo {author} {\bibfnamefont {A.~A.}\ \bibnamefont
  {Saberi}},\ }\bibfield  {title} {\enquote {\bibinfo {title} {Recent advances
  in percolation theory and its applications},}\ }\href {\doibase
  10.1016/j.physrep.2015.03.003} {\bibfield  {journal} {\bibinfo  {journal}
  {Physics Reports}\ }\textbf {\bibinfo {volume} {578}},\ \bibinfo {pages}
  {1--32} (\bibinfo {year} {2015})}\BibitemShut {NoStop}%
\bibitem [{\citenamefont {Stauffer}\ and\ \citenamefont
  {Aharony}(1994)}]{bookDS}%
  \BibitemOpen
  \bibfield  {author} {\bibinfo {author} {\bibfnamefont {D.}~\bibnamefont
  {Stauffer}}\ and\ \bibinfo {author} {\bibfnamefont {A.}~\bibnamefont
  {Aharony}},\ }\href {\doibase 10.1201/9781315274386} {\emph {\bibinfo {title}
  {Introduction to Percolation Theory}}},\ \bibinfo {edition} {2nd}\ ed.\
  (\bibinfo  {publisher} {Taylor and Francis},\ \bibinfo {address} {London},\
  \bibinfo {year} {1994})\BibitemShut {NoStop}%
\bibitem [{\citenamefont {Wierman}(2014)}]{Wierman2014}%
  \BibitemOpen
  \bibfield  {author} {\bibinfo {author} {\bibfnamefont {J.}~\bibnamefont
  {Wierman}},\ }\enquote {\bibinfo {title} {Percolation theory},}\ in\ \href
  {\doibase 10.1002/9781118445112.stat02317} {\emph {\bibinfo {booktitle}
  {Wiley StatsRef: Statistics Reference Online}}}\ (\bibinfo  {publisher}
  {American Cancer Society},\ \bibinfo {year} {2014})\ pp.\ \bibinfo {pages}
  {1--9}\BibitemShut {NoStop}%
\bibitem [{\citenamefont {Dean}(1963)}]{Dean_1963}%
  \BibitemOpen
  \bibfield  {author} {\bibinfo {author} {\bibfnamefont {P.}~\bibnamefont
  {Dean}},\ }\bibfield  {title} {\enquote {\bibinfo {title} {A new {M}onte
  {C}arlo method for percolation problems on a lattice},}\ }\href {\doibase
  10.1017/S0305004100037026} {\bibfield  {journal} {\bibinfo  {journal}
  {Mathematical Proceedings of the Cambridge Philosophical Society}\ }\textbf
  {\bibinfo {volume} {59}},\ \bibinfo {pages} {397--410} (\bibinfo {year}
  {1963})}\BibitemShut {NoStop}%
\bibitem [{\citenamefont {Dean}\ and\ \citenamefont
  {Bird}(1967)}]{Dean_Bird_1967}%
  \BibitemOpen
  \bibfield  {author} {\bibinfo {author} {\bibfnamefont {P.}~\bibnamefont
  {Dean}}\ and\ \bibinfo {author} {\bibfnamefont {N.~F.}\ \bibnamefont
  {Bird}},\ }\bibfield  {title} {\enquote {\bibinfo {title} {Monte {C}arlo
  estimates of critical percolation probabilities},}\ }\href {\doibase
  10.1017/S0305004100041438} {\bibfield  {journal} {\bibinfo  {journal}
  {Mathematical Proceedings of the Cambridge Philosophical Society}\ }\textbf
  {\bibinfo {volume} {63}},\ \bibinfo {pages} {477--479} (\bibinfo {year}
  {1967})}\BibitemShut {NoStop}%
\bibitem [{\citenamefont {Suding}\ and\ \citenamefont
  {Ziff}(1999)}]{PhysRevE.60.275}%
  \BibitemOpen
  \bibfield  {author} {\bibinfo {author} {\bibfnamefont {P.~N.}\ \bibnamefont
  {Suding}}\ and\ \bibinfo {author} {\bibfnamefont {R.~M.}\ \bibnamefont
  {Ziff}},\ }\bibfield  {title} {\enquote {\bibinfo {title} {Site percolation
  thresholds for {A}rchimedean lattices},}\ }\href {\doibase
  10.1103/PhysRevE.60.275} {\bibfield  {journal} {\bibinfo  {journal} {Physical
  Review E}\ }\textbf {\bibinfo {volume} {60}},\ \bibinfo {pages} {275--283}
  (\bibinfo {year} {1999})}\BibitemShut {NoStop}%
\bibitem [{\citenamefont {Dalton}\ \emph {et~al.}(1964)\citenamefont {Dalton},
  \citenamefont {Domb},\ and\ \citenamefont {Sykes}}]{Dalton_1964}%
  \BibitemOpen
  \bibfield  {author} {\bibinfo {author} {\bibfnamefont {N.~W.}\ \bibnamefont
  {Dalton}}, \bibinfo {author} {\bibfnamefont {C.}~\bibnamefont {Domb}}, \ and\
  \bibinfo {author} {\bibfnamefont {M.~F.}\ \bibnamefont {Sykes}},\ }\bibfield
  {title} {\enquote {\bibinfo {title} {Dependence of critical concentration of
  a dilute ferromagnet on the range of interaction},}\ }\href {\doibase
  10.1088/0370-1328/83/3/118} {\bibfield  {journal} {\bibinfo  {journal}
  {Proceedings of the Physical Society}\ }\textbf {\bibinfo {volume} {83}},\
  \bibinfo {pages} {496--498} (\bibinfo {year} {1964})}\BibitemShut {NoStop}%
\bibitem [{\citenamefont {Domb}\ and\ \citenamefont {Dalton}(1966)}]{Domb1966}%
  \BibitemOpen
  \bibfield  {author} {\bibinfo {author} {\bibfnamefont {C.}~\bibnamefont
  {Domb}}\ and\ \bibinfo {author} {\bibfnamefont {N.~W.}\ \bibnamefont
  {Dalton}},\ }\bibfield  {title} {\enquote {\bibinfo {title} {Crystal
  statistics with long-range forces: {I}. {T}he equivalent neighbour model},}\
  }\href {\doibase 10.1088/0370-1328/89/4/311} {\bibfield  {journal} {\bibinfo
  {journal} {Proceedings of the Physical Society}\ }\textbf {\bibinfo {volume}
  {89}},\ \bibinfo {pages} {859--871} (\bibinfo {year} {1966})}\BibitemShut
  {NoStop}%
\bibitem [{\citenamefont {Gouker}\ and\ \citenamefont
  {Family}(1983)}]{Gouker1983}%
  \BibitemOpen
  \bibfield  {author} {\bibinfo {author} {\bibfnamefont {M.}~\bibnamefont
  {Gouker}}\ and\ \bibinfo {author} {\bibfnamefont {F.}~\bibnamefont
  {Family}},\ }\bibfield  {title} {\enquote {\bibinfo {title} {Evidence for
  classical critical behavior in long-range site percolation},}\ }\href
  {\doibase 10.1103/PhysRevB.28.1449} {\bibfield  {journal} {\bibinfo
  {journal} {Physical Review B}\ }\textbf {\bibinfo {volume} {28}},\ \bibinfo
  {pages} {1449--1452} (\bibinfo {year} {1983})}\BibitemShut {NoStop}%
\bibitem [{\citenamefont {Malarz}\ and\ \citenamefont
  {Galam}(2005)}]{Galam2005a}%
  \BibitemOpen
  \bibfield  {author} {\bibinfo {author} {\bibfnamefont {K.}~\bibnamefont
  {Malarz}}\ and\ \bibinfo {author} {\bibfnamefont {S.}~\bibnamefont {Galam}},\
  }\bibfield  {title} {\enquote {\bibinfo {title} {Square-lattice site
  percolation at increasing ranges of neighbor bonds},}\ }\href {\doibase
  10.1103/PhysRevE.71.016125} {\bibfield  {journal} {\bibinfo  {journal}
  {Physical Review E}\ }\textbf {\bibinfo {volume} {71}},\ \bibinfo {pages}
  {016125} (\bibinfo {year} {2005})}\BibitemShut {NoStop}%
\bibitem [{\citenamefont {Galam}\ and\ \citenamefont
  {Malarz}(2005)}]{Galam2005b}%
  \BibitemOpen
  \bibfield  {author} {\bibinfo {author} {\bibfnamefont {S.}~\bibnamefont
  {Galam}}\ and\ \bibinfo {author} {\bibfnamefont {K.}~\bibnamefont {Malarz}},\
  }\bibfield  {title} {\enquote {\bibinfo {title} {Restoring site percolation
  on damaged square lattices},}\ }\href {\doibase 10.1103/PhysRevE.72.027103}
  {\bibfield  {journal} {\bibinfo  {journal} {Physical Review E}\ }\textbf
  {\bibinfo {volume} {72}},\ \bibinfo {pages} {027103} (\bibinfo {year}
  {2005})}\BibitemShut {NoStop}%
\bibitem [{\citenamefont {Majewski}\ and\ \citenamefont
  {Malarz}(2007)}]{Majewski2007}%
  \BibitemOpen
  \bibfield  {author} {\bibinfo {author} {\bibfnamefont {M.}~\bibnamefont
  {Majewski}}\ and\ \bibinfo {author} {\bibfnamefont {K.}~\bibnamefont
  {Malarz}},\ }\bibfield  {title} {\enquote {\bibinfo {title} {Square lattice
  site percolation thresholds for complex neighbourhoods},}\ }\href
  {http://www.actaphys.uj.edu.pl/fulltext?series=Reg&vol=38&page=2191}
  {\bibfield  {journal} {\bibinfo  {journal} {Acta Physica Polonica B}\
  }\textbf {\bibinfo {volume} {38}},\ \bibinfo {pages} {2191--2199} (\bibinfo
  {year} {2007})}\BibitemShut {NoStop}%
\bibitem [{\citenamefont {Xun}\ \emph {et~al.}(2021)\citenamefont {Xun},
  \citenamefont {Hao},\ and\ \citenamefont {Ziff}}]{PhysRevE.103.022126}%
  \BibitemOpen
  \bibfield  {author} {\bibinfo {author} {\bibfnamefont {Z.}~\bibnamefont
  {Xun}}, \bibinfo {author} {\bibfnamefont {D.}~\bibnamefont {Hao}}, \ and\
  \bibinfo {author} {\bibfnamefont {R.~M.}\ \bibnamefont {Ziff}},\ }\bibfield
  {title} {\enquote {\bibinfo {title} {Site percolation on square and simple
  cubic lattices with extended neighborhoods and their continuum limit},}\
  }\href {\doibase 10.1103/PhysRevE.103.022126} {\bibfield  {journal} {\bibinfo
   {journal} {Physical Review E}\ }\textbf {\bibinfo {volume} {103}},\ \bibinfo
  {pages} {022126} (\bibinfo {year} {2021})}\BibitemShut {NoStop}%
\bibitem [{\citenamefont {d'Iribarne}\ \emph {et~al.}(1999)\citenamefont
  {d'Iribarne}, \citenamefont {Rasigni},\ and\ \citenamefont
  {Rasigni}}]{Iribarne1999}%
  \BibitemOpen
  \bibfield  {author} {\bibinfo {author} {\bibfnamefont {C.}~\bibnamefont
  {d'Iribarne}}, \bibinfo {author} {\bibfnamefont {M.}~\bibnamefont {Rasigni}},
  \ and\ \bibinfo {author} {\bibfnamefont {G.}~\bibnamefont {Rasigni}},\
  }\bibfield  {title} {\enquote {\bibinfo {title} {From lattice long-range
  percolation to the continuum one},}\ }\href {\doibase
  10.1016/S0375-9601(99)00585-X} {\bibfield  {journal} {\bibinfo  {journal}
  {Physics Letters A}\ }\textbf {\bibinfo {volume} {263}},\ \bibinfo {pages}
  {65--69} (\bibinfo {year} {1999})}\BibitemShut {NoStop}%
\bibitem [{\citenamefont {Malarz}(2020)}]{2006.15621}%
  \BibitemOpen
  \bibfield  {author} {\bibinfo {author} {\bibfnamefont {K.}~\bibnamefont
  {Malarz}},\ }\bibfield  {title} {\enquote {\bibinfo {title} {Site percolation
  thresholds on triangular lattice with complex neighborhoods},}\ }\href
  {\doibase 10.1063/5.0022336} {\bibfield  {journal} {\bibinfo  {journal}
  {Chaos}\ }\textbf {\bibinfo {volume} {30}},\ \bibinfo {pages} {123123}
  (\bibinfo {year} {2020})}\BibitemShut {NoStop}%
\bibitem [{\citenamefont {Malarz}(2021)}]{2102.10066}%
  \BibitemOpen
  \bibfield  {author} {\bibinfo {author} {\bibfnamefont {K.}~\bibnamefont
  {Malarz}},\ }\bibfield  {title} {\enquote {\bibinfo {title} {Percolation
  thresholds on triangular lattice for neighbourhoods containing sites up-to
  the fifth coordination zone},}\ }\href {\doibase 10.1103/PhysRevE.103.052107}
  {\bibfield  {journal} {\bibinfo  {journal} {Physical Review E}\ }\textbf
  {\bibinfo {volume} {103}},\ \bibinfo {pages} {052107} (\bibinfo {year}
  {2021})}\BibitemShut {NoStop}%
\bibitem [{\citenamefont {Malarz}(2022)}]{2204.12593}%
  \BibitemOpen
  \bibfield  {author} {\bibinfo {author} {\bibfnamefont {K.}~\bibnamefont
  {Malarz}},\ }\bibfield  {title} {\enquote {\bibinfo {title} {Random site
  percolation on honeycomb lattices with complex neighborhoods},}\ }\href
  {\doibase 10.1063/5.0099066} {\bibfield  {journal} {\bibinfo  {journal}
  {Chaos}\ }\textbf {\bibinfo {volume} {32}},\ \bibinfo {pages} {083123}
  (\bibinfo {year} {2022})}\BibitemShut {NoStop}%
\bibitem [{\citenamefont {Lebrecht}\ \emph {et~al.}(2021)\citenamefont
  {Lebrecht}, \citenamefont {Centres},\ and\ \citenamefont
  {Ramirez-Pastor}}]{Lebrecht_2021}%
  \BibitemOpen
  \bibfield  {author} {\bibinfo {author} {\bibfnamefont {W.}~\bibnamefont
  {Lebrecht}}, \bibinfo {author} {\bibfnamefont {P.~M.}\ \bibnamefont
  {Centres}}, \ and\ \bibinfo {author} {\bibfnamefont {A.~J.}\ \bibnamefont
  {Ramirez-Pastor}},\ }\bibfield  {title} {\enquote {\bibinfo {title}
  {Empirical formula for site and bond percolation thresholds on {A}rchimedean
  and 2-uniform lattices},}\ }\href {\doibase 10.1016/j.physa.2021.125802}
  {\bibfield  {journal} {\bibinfo  {journal} {Physica A: Statistical Mechanics
  and its Applications}\ }\textbf {\bibinfo {volume} {569}},\ \bibinfo {pages}
  {125802} (\bibinfo {year} {2021})}\BibitemShut {NoStop}%
\bibitem [{\citenamefont {Xun}\ \emph {et~al.}(2022)\citenamefont {Xun},
  \citenamefont {Hao},\ and\ \citenamefont {Ziff}}]{PhysRevE.105.024105}%
  \BibitemOpen
  \bibfield  {author} {\bibinfo {author} {\bibfnamefont {Z.}~\bibnamefont
  {Xun}}, \bibinfo {author} {\bibfnamefont {D.}~\bibnamefont {Hao}}, \ and\
  \bibinfo {author} {\bibfnamefont {R.~M.}\ \bibnamefont {Ziff}},\ }\bibfield
  {title} {\enquote {\bibinfo {title} {Site and bond percolation thresholds on
  regular lattices with compact extended-range neighborhoods in two and three
  dimensions},}\ }\href {\doibase 10.1103/PhysRevE.105.024105} {\bibfield
  {journal} {\bibinfo  {journal} {Physical Review E}\ }\textbf {\bibinfo
  {volume} {105}},\ \bibinfo {pages} {024105} (\bibinfo {year}
  {2022})}\BibitemShut {NoStop}%
\bibitem [{\citenamefont {Kurzawski}\ and\ \citenamefont
  {Malarz}(2012)}]{Kurzawski2012}%
  \BibitemOpen
  \bibfield  {author} {\bibinfo {author} {\bibfnamefont {{\L}.}~\bibnamefont
  {Kurzawski}}\ and\ \bibinfo {author} {\bibfnamefont {K.}~\bibnamefont
  {Malarz}},\ }\bibfield  {title} {\enquote {\bibinfo {title} {Simple cubic
  random-site percolation thresholds for complex neighbourhoods},}\ }\href
  {\doibase 10.1016/S0034-4877(12)60036-6} {\bibfield  {journal} {\bibinfo
  {journal} {Reports on Mathematical Physics}\ }\textbf {\bibinfo {volume}
  {70}},\ \bibinfo {pages} {163--169} (\bibinfo {year} {2012})}\BibitemShut
  {NoStop}%
\bibitem [{\citenamefont {Malarz}(2015)}]{Malarz2015}%
  \BibitemOpen
  \bibfield  {author} {\bibinfo {author} {\bibfnamefont {K.}~\bibnamefont
  {Malarz}},\ }\bibfield  {title} {\enquote {\bibinfo {title} {Simple cubic
  random-site percolation thresholds for neighborhoods containing
  fourth-nearest neighbors},}\ }\href {\doibase 10.1103/PhysRevE.91.043301}
  {\bibfield  {journal} {\bibinfo  {journal} {Physical Review E}\ }\textbf
  {\bibinfo {volume} {91}},\ \bibinfo {pages} {043301} (\bibinfo {year}
  {2015})}\BibitemShut {NoStop}%
\bibitem [{\citenamefont {Kotwica}\ \emph {et~al.}(2019)\citenamefont
  {Kotwica}, \citenamefont {Gronek},\ and\ \citenamefont
  {Malarz}}]{1803.09504}%
  \BibitemOpen
  \bibfield  {author} {\bibinfo {author} {\bibfnamefont {M.}~\bibnamefont
  {Kotwica}}, \bibinfo {author} {\bibfnamefont {P.}~\bibnamefont {Gronek}}, \
  and\ \bibinfo {author} {\bibfnamefont {K.}~\bibnamefont {Malarz}},\
  }\bibfield  {title} {\enquote {\bibinfo {title} {Efficient space
  virtualisation for {H}oshen--{K}opelman algorithm},}\ }\href {\doibase
  10.1142/S0129183119500554} {\bibfield  {journal} {\bibinfo  {journal}
  {International Journal of Modern Physics C}\ }\textbf {\bibinfo {volume}
  {30}},\ \bibinfo {pages} {1950055} (\bibinfo {year} {2019})}\BibitemShut
  {NoStop}%
\bibitem [{\citenamefont {Zhao}\ \emph {et~al.}(2022)\citenamefont {Zhao},
  \citenamefont {Yan}, \citenamefont {Xun}, \citenamefont {Hao},\ and\
  \citenamefont {Ziff}}]{Zhao_2022}%
  \BibitemOpen
  \bibfield  {author} {\bibinfo {author} {\bibfnamefont {P.}~\bibnamefont
  {Zhao}}, \bibinfo {author} {\bibfnamefont {J.}~\bibnamefont {Yan}}, \bibinfo
  {author} {\bibfnamefont {Z.}~\bibnamefont {Xun}}, \bibinfo {author}
  {\bibfnamefont {D.}~\bibnamefont {Hao}}, \ and\ \bibinfo {author}
  {\bibfnamefont {R.~M.}\ \bibnamefont {Ziff}},\ }\bibfield  {title} {\enquote
  {\bibinfo {title} {Site and bond percolation on four-dimensional simple
  hypercubic lattices with extended neighborhoods},}\ }\href {\doibase
  10.1088/1742-5468/ac52a8} {\bibfield  {journal} {\bibinfo  {journal} {Journal
  of Statistical Mechanics: Theory and Experiment}\ }\textbf {\bibinfo {volume}
  {2022}},\ \bibinfo {pages} {033202} (\bibinfo {year} {2022})}\BibitemShut
  {NoStop}%
\bibitem [{\citenamefont {Xu}\ \emph {et~al.}(2021)\citenamefont {Xu},
  \citenamefont {Wang}, \citenamefont {Hu},\ and\ \citenamefont
  {Deng}}]{PhysRevE.103.022127}%
  \BibitemOpen
  \bibfield  {author} {\bibinfo {author} {\bibfnamefont {W.}~\bibnamefont
  {Xu}}, \bibinfo {author} {\bibfnamefont {J.}~\bibnamefont {Wang}}, \bibinfo
  {author} {\bibfnamefont {H.}~\bibnamefont {Hu}}, \ and\ \bibinfo {author}
  {\bibfnamefont {Y.}~\bibnamefont {Deng}},\ }\bibfield  {title} {\enquote
  {\bibinfo {title} {Critical polynomials in the nonplanar and continuum
  percolation models},}\ }\href {\doibase 10.1103/PhysRevE.103.022127}
  {\bibfield  {journal} {\bibinfo  {journal} {Physical Review E}\ }\textbf
  {\bibinfo {volume} {103}},\ \bibinfo {pages} {022127} (\bibinfo {year}
  {2021})}\BibitemShut {NoStop}%
\bibitem [{\citenamefont {Koza}\ \emph {et~al.}(2014)\citenamefont {Koza},
  \citenamefont {Kondrat},\ and\ \citenamefont
  {Suszczy{\'{n}}ski}}]{Koza_2014}%
  \BibitemOpen
  \bibfield  {author} {\bibinfo {author} {\bibfnamefont {Z.}~\bibnamefont
  {Koza}}, \bibinfo {author} {\bibfnamefont {G.}~\bibnamefont {Kondrat}}, \
  and\ \bibinfo {author} {\bibfnamefont {K.}~\bibnamefont
  {Suszczy{\'{n}}ski}},\ }\bibfield  {title} {\enquote {\bibinfo {title}
  {Percolation of overlapping squares or cubes on a lattice},}\ }\href
  {\doibase 10.1088/1742-5468/2014/11/p11005} {\bibfield  {journal} {\bibinfo
  {journal} {Journal of Statistical Mechanics: Theory and Experiment}\ }\textbf
  {\bibinfo {volume} {2014}},\ \bibinfo {pages} {P11005} (\bibinfo {year}
  {2014})}\BibitemShut {NoStop}%
\bibitem [{\citenamefont {Koza}\ and\ \citenamefont
  {Po{\l}a}(2016)}]{Koza_2016}%
  \BibitemOpen
  \bibfield  {author} {\bibinfo {author} {\bibfnamefont {Z.}~\bibnamefont
  {Koza}}\ and\ \bibinfo {author} {\bibfnamefont {J.}~\bibnamefont {Po{\l}a}},\
  }\bibfield  {title} {\enquote {\bibinfo {title} {From discrete to continuous
  percolation in dimensions 3 to 7},}\ }\href {\doibase
  10.1088/1742-5468/2016/10/103206} {\bibfield  {journal} {\bibinfo  {journal}
  {Journal of Statistical Mechanics: Theory and Experiment}\ }\textbf {\bibinfo
  {volume} {2016}},\ \bibinfo {pages} {103206} (\bibinfo {year}
  {2016})}\BibitemShut {NoStop}%
\bibitem [{\citenamefont {Galam}\ and\ \citenamefont
  {Mauger}(1996)}]{PhysRevE.53.2177}%
  \BibitemOpen
  \bibfield  {author} {\bibinfo {author} {\bibfnamefont {S.}~\bibnamefont
  {Galam}}\ and\ \bibinfo {author} {\bibfnamefont {A.}~\bibnamefont {Mauger}},\
  }\bibfield  {title} {\enquote {\bibinfo {title} {Universal formulas for
  percolation thresholds},}\ }\href {\doibase 10.1103/PhysRevE.53.2177}
  {\bibfield  {journal} {\bibinfo  {journal} {Physical Review E}\ }\textbf
  {\bibinfo {volume} {53}},\ \bibinfo {pages} {2177--2181} (\bibinfo {year}
  {1996})}\BibitemShut {NoStop}%
\bibitem [{\citenamefont {Galam}\ and\ \citenamefont
  {Mauger}(1997)}]{PhysRevE.55.1230}%
  \BibitemOpen
  \bibfield  {author} {\bibinfo {author} {\bibfnamefont {S.}~\bibnamefont
  {Galam}}\ and\ \bibinfo {author} {\bibfnamefont {A.}~\bibnamefont {Mauger}},\
  }\bibfield  {title} {\enquote {\bibinfo {title} {Reply to ``{C}omment on
  `{U}niversal formulas for percolation thresholds'{''}},}\ }\href {\doibase
  10.1103/PhysRevE.55.1230} {\bibfield  {journal} {\bibinfo  {journal}
  {Physical Review E}\ }\textbf {\bibinfo {volume} {55}},\ \bibinfo {pages}
  {1230--1231} (\bibinfo {year} {1997})}\BibitemShut {NoStop}%
\bibitem [{\citenamefont {van~der Marck}(1997)}]{ISI:A1997WD54600080}%
  \BibitemOpen
  \bibfield  {author} {\bibinfo {author} {\bibfnamefont {S.~C.}\ \bibnamefont
  {van~der Marck}},\ }\bibfield  {title} {\enquote {\bibinfo {title} {Universal
  formulas for percolation thresholds --- {C}omment},}\ }\href {\doibase
  10.1103/PhysRevE.55.1228} {\bibfield  {journal} {\bibinfo  {journal}
  {Physical Review E}\ }\textbf {\bibinfo {volume} {55}},\ \bibinfo {pages}
  {1228--1229} (\bibinfo {year} {1997})}\BibitemShut {NoStop}%
\bibitem [{\citenamefont {Newman}\ and\ \citenamefont
  {Ziff}(2001)}]{NewmanZiff2001}%
  \BibitemOpen
  \bibfield  {author} {\bibinfo {author} {\bibfnamefont {M.~E.~J.}\
  \bibnamefont {Newman}}\ and\ \bibinfo {author} {\bibfnamefont {R.~M.}\
  \bibnamefont {Ziff}},\ }\bibfield  {title} {\enquote {\bibinfo {title} {Fast
  {M}onte {C}arlo algorithm for site or bond percolation},}\ }\href {\doibase
  10.1103/PhysRevE.64.016706} {\bibfield  {journal} {\bibinfo  {journal}
  {Physical Review E}\ }\textbf {\bibinfo {volume} {64}},\ \bibinfo {pages}
  {016706} (\bibinfo {year} {2001})}\BibitemShut {NoStop}%
\bibitem [{\citenamefont {Privman}(1990)}]{Finite-Size_Scaling_Theory_1990}%
  \BibitemOpen
  \bibfield  {author} {\bibinfo {author} {\bibfnamefont {V.}~\bibnamefont
  {Privman}},\ }\enquote {\bibinfo {title} {Finite-size scaling theory},}\ in\
  \href {\doibase 10.1142/9789814503419_0001} {\emph {\bibinfo {booktitle}
  {Finite size scaling and numerical simulation of statistical systems}}},\
  \bibinfo {editor} {edited by\ \bibinfo {editor} {\bibfnamefont
  {V.}~\bibnamefont {Privman}}}\ (\bibinfo  {publisher} {World Scientific},\
  \bibinfo {address} {Singapore},\ \bibinfo {year} {1990})\ pp.\ \bibinfo
  {pages} {1--98}\BibitemShut {NoStop}%
\bibitem [{\citenamefont {Landau}\ and\ \citenamefont
  {Binder}(2009)}]{Guide_to_Monte_Carlo_Simulations_2009}%
  \BibitemOpen
  \bibfield  {author} {\bibinfo {author} {\bibfnamefont {D.~P.}\ \bibnamefont
  {Landau}}\ and\ \bibinfo {author} {\bibfnamefont {K.}~\bibnamefont
  {Binder}},\ }\href {\doibase 10.1017/CBO9780511994944} {\emph {\bibinfo
  {title} {A Guide to Monte Carlo Simulations in Statistical Physics}}},\
  \bibinfo {edition} {3rd}\ ed.\ (\bibinfo  {publisher} {Cambridge University
  Press},\ \bibinfo {year} {2009})\BibitemShut {NoStop}%
\bibitem [{\citenamefont {Newman}\ and\ \citenamefont
  {Ziff}(2000)}]{PhysRevLett.85.4104}%
  \BibitemOpen
  \bibfield  {author} {\bibinfo {author} {\bibfnamefont {M.~E.~J.}\
  \bibnamefont {Newman}}\ and\ \bibinfo {author} {\bibfnamefont {R.~M.}\
  \bibnamefont {Ziff}},\ }\bibfield  {title} {\enquote {\bibinfo {title}
  {Efficient {M}onte {C}arlo algorithm and high-precision results for
  percolation},}\ }\href {\doibase 10.1103/PhysRevLett.85.4104} {\bibfield
  {journal} {\bibinfo  {journal} {Physical Review Letters}\ }\textbf {\bibinfo
  {volume} {85}},\ \bibinfo {pages} {4104--4107} (\bibinfo {year}
  {2000})}\BibitemShut {NoStop}%
\bibitem [{\citenamefont {Tencer}\ and\ \citenamefont
  {Forsberg}(2021)}]{PhysRevE.103.012115}%
  \BibitemOpen
  \bibfield  {author} {\bibinfo {author} {\bibfnamefont {J.}~\bibnamefont
  {Tencer}}\ and\ \bibinfo {author} {\bibfnamefont {K.~M.}\ \bibnamefont
  {Forsberg}},\ }\bibfield  {title} {\enquote {\bibinfo {title} {Postprocessing
  techniques for gradient percolation predictions on the square lattice},}\
  }\href {\doibase 10.1103/PhysRevE.103.012115} {\bibfield  {journal} {\bibinfo
   {journal} {Physical Review E}\ }\textbf {\bibinfo {volume} {103}},\ \bibinfo
  {pages} {012115} (\bibinfo {year} {2021})}\BibitemShut {NoStop}%
\bibitem [{\citenamefont {Xun}\ and\ \citenamefont {Hao}(2022)}]{Xun_2022}%
  \BibitemOpen
  \bibfield  {author} {\bibinfo {author} {\bibfnamefont {Z.}~\bibnamefont
  {Xun}}\ and\ \bibinfo {author} {\bibfnamefont {D.}~\bibnamefont {Hao}},\
  }\bibfield  {title} {\enquote {\bibinfo {title} {Monte {C}arlo simulation of
  bond percolation on square lattice with complex neighborhoods},}\ }\href
  {\doibase 10.7498/aps.71.20211757} {\bibfield  {journal} {\bibinfo  {journal}
  {Acta Physica Sinica}\ }\textbf {\bibinfo {volume} {71}},\ \bibinfo {pages}
  {066401} (\bibinfo {year} {2022})},\ \bibinfo {note} {in Chinese}\BibitemShut
  {NoStop}%
\bibitem [{\citenamefont {Cruz}\ \emph {et~al.}(2023)\citenamefont {Cruz},
  \citenamefont {Ortiz}, \citenamefont {Ortiz},\ and\ \citenamefont
  {Balankin}}]{fractalfract7030231}%
  \BibitemOpen
  \bibfield  {author} {\bibinfo {author} {\bibfnamefont {M.-\'A.~M.}\
  \bibnamefont {Cruz}}, \bibinfo {author} {\bibfnamefont {J.~P.}\ \bibnamefont
  {Ortiz}}, \bibinfo {author} {\bibfnamefont {M.~P.}\ \bibnamefont {Ortiz}}, \
  and\ \bibinfo {author} {\bibfnamefont {A.}~\bibnamefont {Balankin}},\
  }\bibfield  {title} {\enquote {\bibinfo {title} {Percolation on fractal
  networks: {A} survey},}\ }\href {\doibase 10.3390/fractalfract7030231}
  {\bibfield  {journal} {\bibinfo  {journal} {Fractal and Fractional}\ }\textbf
  {\bibinfo {volume} {7}},\ \bibinfo {pages} {231} (\bibinfo {year}
  {2023})}\BibitemShut {NoStop}%
\bibitem [{\citenamefont {Galam}\ and\ \citenamefont
  {Mauger}(1994)}]{Galam_Mauger_1994}%
  \BibitemOpen
  \bibfield  {author} {\bibinfo {author} {\bibfnamefont {S.}~\bibnamefont
  {Galam}}\ and\ \bibinfo {author} {\bibfnamefont {A.}~\bibnamefont {Mauger}},\
  }\bibfield  {title} {\enquote {\bibinfo {title} {A new scheme to percolation
  thresholds},}\ }\href {\doibase 10.1063/1.355677} {\bibfield  {journal}
  {\bibinfo  {journal} {J. Appl. Phys.}\ }\textbf {\bibinfo {volume} {75}},\
  \bibinfo {pages} {5526--5528} (\bibinfo {year} {1994})}\BibitemShut {NoStop}%
\bibitem [{\citenamefont {Mertens}\ and\ \citenamefont
  {Moore}(2018)}]{PhysRevE.98.022120}%
  \BibitemOpen
  \bibfield  {author} {\bibinfo {author} {\bibfnamefont {S.}~\bibnamefont
  {Mertens}}\ and\ \bibinfo {author} {\bibfnamefont {C.}~\bibnamefont
  {Moore}},\ }\bibfield  {title} {\enquote {\bibinfo {title} {Percolation
  thresholds and {F}isher exponents in hypercubic lattices},}\ }\href {\doibase
  10.1103/PhysRevE.98.022120} {\bibfield  {journal} {\bibinfo  {journal}
  {Physical Review E}\ }\textbf {\bibinfo {volume} {98}},\ \bibinfo {pages}
  {022120} (\bibinfo {year} {2018})}\BibitemShut {NoStop}%
\bibitem [{\citenamefont {Xun}\ \emph {et~al.}(2023)\citenamefont {Xun},
  \citenamefont {Hao},\ and\ \citenamefont {Ziff}}]{2308.15719}%
  \BibitemOpen
  \bibfield  {author} {\bibinfo {author} {\bibfnamefont {Z.}~\bibnamefont
  {Xun}}, \bibinfo {author} {\bibfnamefont {D.}~\bibnamefont {Hao}}, \ and\
  \bibinfo {author} {\bibfnamefont {R.~M.}\ \bibnamefont {Ziff}},\ }\href@noop
  {} {\enquote {\bibinfo {title} {Extended-range percolation in five
  dimensions},}\ } (\bibinfo {year} {2023}),\ \Eprint
  {http://arxiv.org/abs/2308.15719} {arXiv:2308.15719 [cond-mat.stat-mech]}
  \BibitemShut {NoStop}%
\end{thebibliography}%

\begin{figure*}[htbp]
\begin{subfigure}[b]{0.14\textwidth}
\caption{\label{fig:16nn}}
\begin{tikzpicture}[scale=0.36]
\draw[step=1,color=gray] (-3.5,-3.5) grid (3.5,3.5);
\draw[very thick,red] (0, 0) circle (.4);
\filldraw[black] (0,-1) circle (.4);
\filldraw[black] (0,+1) circle (.4);
\filldraw[black] (-1,0) circle (.4);
\filldraw[black] (+1,0) circle (.4);
\filldraw[black] (0,-3) circle (.4);
\filldraw[black] (0,+3) circle (.4);
\filldraw[black] (-3,0) circle (.4);
\filldraw[black] (+3,0) circle (.4);
\end{tikzpicture}
\end{subfigure}
\hfill
\begin{subfigure}[b]{0.14\textwidth}
\caption{\label{fig:26nn}}
\begin{tikzpicture}[scale=0.36]
\draw[step=1,color=gray] (-3.5,-3.5) grid (3.5,3.5);
\draw[very thick,red] (0, 0) circle (.4);
\filldraw[black] (+1,-1) circle (.4);
\filldraw[black] (+1,+1) circle (.4);
\filldraw[black] (-1,-1) circle (.4);
\filldraw[black] (-1,+1) circle (.4);
\filldraw[black] (0,-3) circle (.4);
\filldraw[black] (0,+3) circle (.4);
\filldraw[black] (-3,0) circle (.4);
\filldraw[black] (+3,0) circle (.4);
\end{tikzpicture}
\end{subfigure}
\hfill
\begin{subfigure}[b]{0.14\textwidth}
\caption{\label{fig:36nn}}
\begin{tikzpicture}[scale=0.36]
\draw[step=1,color=gray] (-3.5,-3.5) grid (3.5,3.5);
\draw[very thick,red] (0, 0) circle (.4);
\filldraw[black] ( 0,-2) circle (.4);
\filldraw[black] ( 0,+2) circle (.4);
\filldraw[black] (-2, 0) circle (.4);
\filldraw[black] (+2, 0) circle (.4);
\filldraw[black] (0,-3) circle (.4);
\filldraw[black] (0,+3) circle (.4);
\filldraw[black] (-3,0) circle (.4);
\filldraw[black] (+3,0) circle (.4);
\end{tikzpicture}
\end{subfigure}
\hfill
\begin{subfigure}[b]{0.14\textwidth}
\caption{\label{fig:46nn}}
\begin{tikzpicture}[scale=0.36]
\draw[step=1,color=gray] (-3.5,-3.5) grid (3.5,3.5);
\draw[very thick,red] (0, 0) circle (.4);
\filldraw[black] (-2,-1) circle (.4);
\filldraw[black] (-2,+1) circle (.4);
\filldraw[black] (-1,-2) circle (.4);
\filldraw[black] (-1,+2) circle (.4);
\filldraw[black] (+1,-2) circle (.4);
\filldraw[black] (+1,+2) circle (.4);
\filldraw[black] (+2,-1) circle (.4);
\filldraw[black] (+2,+1) circle (.4);
\filldraw[black] (0,-3) circle (.4);
\filldraw[black] (0,+3) circle (.4);
\filldraw[black] (-3,0) circle (.4);
\filldraw[black] (+3,0) circle (.4);
\end{tikzpicture}
\end{subfigure}
\hfill
\begin{subfigure}[b]{0.14\textwidth}
\caption{\label{fig:56nn}}
\begin{tikzpicture}[scale=0.36]
\draw[step=1,color=gray] (-3.5,-3.5) grid (3.5,3.5);
\draw[very thick,red] (0, 0) circle (.4);
\filldraw[black] (-2,-2) circle (.4);
\filldraw[black] (-2,+2) circle (.4);
\filldraw[black] (+2,-2) circle (.4);
\filldraw[black] (+2,+2) circle (.4);
\filldraw[black] (0,-3) circle (.4);
\filldraw[black] (0,+3) circle (.4);
\filldraw[black] (-3,0) circle (.4);
\filldraw[black] (+3,0) circle (.4);
\end{tikzpicture}
\end{subfigure}
\hfill
\begin{subfigure}[b]{0.14\textwidth}
\caption{\label{fig:126nn}}
\begin{tikzpicture}[scale=0.36]
\draw[step=1,color=gray] (-3.5,-3.5) grid (3.5,3.5);
\draw[very thick,red] (0, 0) circle (.4);
\filldraw[black] (0,-1) circle (.4);
\filldraw[black] (0,+1) circle (.4);
\filldraw[black] (-1,0) circle (.4);
\filldraw[black] (+1,0) circle (.4);
\filldraw[black] (+1,-1) circle (.4);
\filldraw[black] (+1,+1) circle (.4);
\filldraw[black] (-1,-1) circle (.4);
\filldraw[black] (-1,+1) circle (.4);
\filldraw[black] (0,-3) circle (.4);
\filldraw[black] (0,+3) circle (.4);
\filldraw[black] (-3,0) circle (.4);
\filldraw[black] (+3,0) circle (.4);
\end{tikzpicture}
\end{subfigure}
\hfill
\begin{subfigure}[b]{0.14\textwidth}
\caption{\label{fig:136nn}}
\begin{tikzpicture}[scale=0.36]
\draw[step=1,color=gray] (-3.5,-3.5) grid (3.5,3.5);
\draw[very thick,red] (0, 0) circle (.4);
\filldraw[black] (0,-1) circle (.4);
\filldraw[black] (0,+1) circle (.4);
\filldraw[black] (-1,0) circle (.4);
\filldraw[black] (+1,0) circle (.4);
\filldraw[black] ( 0,-2) circle (.4);
\filldraw[black] ( 0,+2) circle (.4);
\filldraw[black] (-2, 0) circle (.4);
\filldraw[black] (+2, 0) circle (.4);
\filldraw[black] (0,-3) circle (.4);
\filldraw[black] (0,+3) circle (.4);
\filldraw[black] (-3,0) circle (.4);
\filldraw[black] (+3,0) circle (.4);
\end{tikzpicture}
\end{subfigure}
\hfill
\begin{subfigure}[b]{0.14\textwidth}
\caption{\label{fig:146nn}}
\begin{tikzpicture}[scale=0.36]
\draw[step=1,color=gray] (-3.5,-3.5) grid (3.5,3.5);
\draw[very thick,red] (0, 0) circle (.4);
\filldraw[black] (0,-1) circle (.4);
\filldraw[black] (0,+1) circle (.4);
\filldraw[black] (-1,0) circle (.4);
\filldraw[black] (+1,0) circle (.4);
\filldraw[black] (-2,-1) circle (.4);
\filldraw[black] (-2,+1) circle (.4);
\filldraw[black] (-1,-2) circle (.4);
\filldraw[black] (-1,+2) circle (.4);
\filldraw[black] (+1,-2) circle (.4);
\filldraw[black] (+1,+2) circle (.4);
\filldraw[black] (+2,-1) circle (.4);
\filldraw[black] (+2,+1) circle (.4);
\filldraw[black] (0,-3) circle (.4);
\filldraw[black] (0,+3) circle (.4);
\filldraw[black] (-3,0) circle (.4);
\filldraw[black] (+3,0) circle (.4);
\end{tikzpicture}
\end{subfigure}
\hfill
\begin{subfigure}[b]{0.14\textwidth}
\caption{\label{fig:156nn}}
\begin{tikzpicture}[scale=0.36]
\draw[step=1,color=gray] (-3.5,-3.5) grid (3.5,3.5);
\draw[very thick,red] (0, 0) circle (.4);
\filldraw[black] (0,-1) circle (.4);
\filldraw[black] (0,+1) circle (.4);
\filldraw[black] (-1,0) circle (.4);
\filldraw[black] (+1,0) circle (.4);
\filldraw[black] (-2,-2) circle (.4);
\filldraw[black] (-2,+2) circle (.4);
\filldraw[black] (+2,-2) circle (.4);
\filldraw[black] (+2,+2) circle (.4);
\filldraw[black] (0,-3) circle (.4);
\filldraw[black] (0,+3) circle (.4);
\filldraw[black] (-3,0) circle (.4);
\filldraw[black] (+3,0) circle (.4);
\end{tikzpicture}
\end{subfigure}
\hfill
\begin{subfigure}[b]{0.14\textwidth}
\caption{\label{fig:236nn}}
\begin{tikzpicture}[scale=0.36]
\draw[step=1,color=gray] (-3.5,-3.5) grid (3.5,3.5);
\draw[very thick,red] (0, 0) circle (.4);
\filldraw[black] (+1,-1) circle (.4);
\filldraw[black] (+1,+1) circle (.4);
\filldraw[black] (-1,-1) circle (.4);
\filldraw[black] (-1,+1) circle (.4);
\filldraw[black] ( 0,-2) circle (.4);
\filldraw[black] ( 0,+2) circle (.4);
\filldraw[black] (-2, 0) circle (.4);
\filldraw[black] (+2, 0) circle (.4);
\filldraw[black] (0,-3) circle (.4);
\filldraw[black] (0,+3) circle (.4);
\filldraw[black] (-3,0) circle (.4);
\filldraw[black] (+3,0) circle (.4);
\end{tikzpicture}
\end{subfigure}
\hfill
\begin{subfigure}[b]{0.14\textwidth}
\caption{\label{fig:246nn}}
\begin{tikzpicture}[scale=0.36]
\draw[step=1,color=gray] (-3.5,-3.5) grid (3.5,3.5);
\draw[very thick,red] (0, 0) circle (.4);
\filldraw[black] (+1,-1) circle (.4);
\filldraw[black] (+1,+1) circle (.4);
\filldraw[black] (-1,-1) circle (.4);
\filldraw[black] (-1,+1) circle (.4);
\filldraw[black] (-2,-1) circle (.4);
\filldraw[black] (-2,+1) circle (.4);
\filldraw[black] (-1,-2) circle (.4);
\filldraw[black] (-1,+2) circle (.4);
\filldraw[black] (+1,-2) circle (.4);
\filldraw[black] (+1,+2) circle (.4);
\filldraw[black] (+2,-1) circle (.4);
\filldraw[black] (+2,+1) circle (.4);
\filldraw[black] (0,-3) circle (.4);
\filldraw[black] (0,+3) circle (.4);
\filldraw[black] (-3,0) circle (.4);
\filldraw[black] (+3,0) circle (.4);
\end{tikzpicture}
\end{subfigure}
\hfill
\begin{subfigure}[b]{0.14\textwidth}
\caption{\label{fig:256nn}}
\begin{tikzpicture}[scale=0.36]
\draw[step=1,color=gray] (-3.5,-3.5) grid (3.5,3.5);
\draw[very thick,red] (0, 0) circle (.4);
\filldraw[black] (+1,-1) circle (.4);
\filldraw[black] (+1,+1) circle (.4);
\filldraw[black] (-1,-1) circle (.4);
\filldraw[black] (-1,+1) circle (.4);
\filldraw[black] (-2,-2) circle (.4);
\filldraw[black] (-2,+2) circle (.4);
\filldraw[black] (+2,-2) circle (.4);
\filldraw[black] (+2,+2) circle (.4);
\filldraw[black] (0,-3) circle (.4);
\filldraw[black] (0,+3) circle (.4);
\filldraw[black] (-3,0) circle (.4);
\filldraw[black] (+3,0) circle (.4);
\end{tikzpicture}
\end{subfigure}
\hfill
\begin{subfigure}[b]{0.14\textwidth}
\caption{\label{fig:346nn}}
\begin{tikzpicture}[scale=0.36]
\draw[step=1,color=gray] (-3.5,-3.5) grid (3.5,3.5);
\draw[very thick,red] (0, 0) circle (.4);
\filldraw[black] ( 0,-2) circle (.4);
\filldraw[black] ( 0,+2) circle (.4);
\filldraw[black] (-2, 0) circle (.4);
\filldraw[black] (+2, 0) circle (.4);
\filldraw[black] (-2,-1) circle (.4);
\filldraw[black] (-2,+1) circle (.4);
\filldraw[black] (-1,-2) circle (.4);
\filldraw[black] (-1,+2) circle (.4);
\filldraw[black] (+1,-2) circle (.4);
\filldraw[black] (+1,+2) circle (.4);
\filldraw[black] (+2,-1) circle (.4);
\filldraw[black] (+2,+1) circle (.4);
\filldraw[black] (0,-3) circle (.4);
\filldraw[black] (0,+3) circle (.4);
\filldraw[black] (-3,0) circle (.4);
\filldraw[black] (+3,0) circle (.4);
\end{tikzpicture}
\end{subfigure}
\hfill
\begin{subfigure}[b]{0.14\textwidth}
\caption{\label{fig:356nn}}
\begin{tikzpicture}[scale=0.36]
\draw[step=1,color=gray] (-3.5,-3.5) grid (3.5,3.5);
\draw[very thick,red] (0, 0) circle (.4);
\filldraw[black] ( 0,-2) circle (.4);
\filldraw[black] ( 0,+2) circle (.4);
\filldraw[black] (-2, 0) circle (.4);
\filldraw[black] (+2, 0) circle (.4);
\filldraw[black] (-2,-2) circle (.4);
\filldraw[black] (-2,+2) circle (.4);
\filldraw[black] (+2,-2) circle (.4);
\filldraw[black] (+2,+2) circle (.4);
\filldraw[black] (0,-3) circle (.4);
\filldraw[black] (0,+3) circle (.4);
\filldraw[black] (-3,0) circle (.4);
\filldraw[black] (+3,0) circle (.4);
\end{tikzpicture}
\end{subfigure}
\hfill
\begin{subfigure}[b]{0.14\textwidth}
\caption{\label{fig:456nn}}
\begin{tikzpicture}[scale=0.36]
\draw[step=1,color=gray] (-3.5,-3.5) grid (3.5,3.5);
\draw[very thick,red] (0, 0) circle (.4);
\filldraw[black] (-2,-1) circle (.4);
\filldraw[black] (-2,+1) circle (.4);
\filldraw[black] (-1,-2) circle (.4);
\filldraw[black] (-1,+2) circle (.4);
\filldraw[black] (+1,-2) circle (.4);
\filldraw[black] (+1,+2) circle (.4);
\filldraw[black] (+2,-1) circle (.4);
\filldraw[black] (+2,+1) circle (.4);
\filldraw[black] (-2,-2) circle (.4);
\filldraw[black] (-2,+2) circle (.4);
\filldraw[black] (+2,-2) circle (.4);
\filldraw[black] (+2,+2) circle (.4);
\filldraw[black] (0,-3) circle (.4);
\filldraw[black] (0,+3) circle (.4);
\filldraw[black] (-3,0) circle (.4);
\filldraw[black] (+3,0) circle (.4);
\end{tikzpicture}
\end{subfigure}
\hfill
\begin{subfigure}[b]{0.14\textwidth}
\caption{\label{fig:1236nn}}
\begin{tikzpicture}[scale=0.36]
\draw[step=1,color=gray] (-3.5,-3.5) grid (3.5,3.5);
\draw[very thick,red] (0, 0) circle (.4);
\filldraw[black] (0,-1) circle (.4);
\filldraw[black] (0,+1) circle (.4);
\filldraw[black] (-1,0) circle (.4);
\filldraw[black] (+1,0) circle (.4);
\filldraw[black] (+1,-1) circle (.4);
\filldraw[black] (+1,+1) circle (.4);
\filldraw[black] (-1,-1) circle (.4);
\filldraw[black] (-1,+1) circle (.4);
\filldraw[black] ( 0,-2) circle (.4);
\filldraw[black] ( 0,+2) circle (.4);
\filldraw[black] (-2, 0) circle (.4);
\filldraw[black] (+2, 0) circle (.4);
\filldraw[black] (0,-3) circle (.4);
\filldraw[black] (0,+3) circle (.4);
\filldraw[black] (-3,0) circle (.4);
\filldraw[black] (+3,0) circle (.4);
\end{tikzpicture}
\end{subfigure}
\hfill
\begin{subfigure}[b]{0.14\textwidth}
\caption{\label{fig:1246nn}}
\begin{tikzpicture}[scale=0.36]
\draw[step=1,color=gray] (-3.5,-3.5) grid (3.5,3.5);
\draw[very thick,red] (0, 0) circle (.4);
\filldraw[black] (0,-1) circle (.4);
\filldraw[black] (0,+1) circle (.4);
\filldraw[black] (-1,0) circle (.4);
\filldraw[black] (+1,0) circle (.4);
\filldraw[black] (+1,-1) circle (.4);
\filldraw[black] (+1,+1) circle (.4);
\filldraw[black] (-1,-1) circle (.4);
\filldraw[black] (-1,+1) circle (.4);
\filldraw[black] (-2,-1) circle (.4);
\filldraw[black] (-2,+1) circle (.4);
\filldraw[black] (-1,-2) circle (.4);
\filldraw[black] (-1,+2) circle (.4);
\filldraw[black] (+1,-2) circle (.4);
\filldraw[black] (+1,+2) circle (.4);
\filldraw[black] (+2,-1) circle (.4);
\filldraw[black] (+2,+1) circle (.4);
\filldraw[black] (0,-3) circle (.4);
\filldraw[black] (0,+3) circle (.4);
\filldraw[black] (-3,0) circle (.4);
\filldraw[black] (+3,0) circle (.4);
\end{tikzpicture}
\end{subfigure}
\hfill
\begin{subfigure}[b]{0.14\textwidth}
\caption{\label{fig:1256nn}}
\begin{tikzpicture}[scale=0.36]
\draw[step=1,color=gray] (-3.5,-3.5) grid (3.5,3.5);
\draw[very thick,red] (0, 0) circle (.4);
\filldraw[black] (0,-1) circle (.4);
\filldraw[black] (0,+1) circle (.4);
\filldraw[black] (-1,0) circle (.4);
\filldraw[black] (+1,0) circle (.4);
\filldraw[black] (+1,-1) circle (.4);
\filldraw[black] (+1,+1) circle (.4);
\filldraw[black] (-1,-1) circle (.4);
\filldraw[black] (-1,+1) circle (.4);
\filldraw[black] (-2,-2) circle (.4);
\filldraw[black] (-2,+2) circle (.4);
\filldraw[black] (+2,-2) circle (.4);
\filldraw[black] (+2,+2) circle (.4);
\filldraw[black] (0,-3) circle (.4);
\filldraw[black] (0,+3) circle (.4);
\filldraw[black] (-3,0) circle (.4);
\filldraw[black] (+3,0) circle (.4);
\end{tikzpicture}
\end{subfigure}
\hfill
\begin{subfigure}[b]{0.14\textwidth}
\caption{\label{fig:1346nn}}
\begin{tikzpicture}[scale=0.36]
\draw[step=1,color=gray] (-3.5,-3.5) grid (3.5,3.5);
\draw[very thick,red] (0, 0) circle (.4);
\filldraw[black] (0,-1) circle (.4);
\filldraw[black] (0,+1) circle (.4);
\filldraw[black] (-1,0) circle (.4);
\filldraw[black] (+1,0) circle (.4);
\filldraw[black] ( 0,-2) circle (.4);
\filldraw[black] ( 0,+2) circle (.4);
\filldraw[black] (-2, 0) circle (.4);
\filldraw[black] (+2, 0) circle (.4);
\filldraw[black] (-2,-1) circle (.4);
\filldraw[black] (-2,+1) circle (.4);
\filldraw[black] (-1,-2) circle (.4);
\filldraw[black] (-1,+2) circle (.4);
\filldraw[black] (+1,-2) circle (.4);
\filldraw[black] (+1,+2) circle (.4);
\filldraw[black] (+2,-1) circle (.4);
\filldraw[black] (+2,+1) circle (.4);
\filldraw[black] (0,-3) circle (.4);
\filldraw[black] (0,+3) circle (.4);
\filldraw[black] (-3,0) circle (.4);
\filldraw[black] (+3,0) circle (.4);
\end{tikzpicture}
\end{subfigure}
\hfill
\begin{subfigure}[b]{0.14\textwidth}
\caption{\label{fig:1356nn}}
\begin{tikzpicture}[scale=0.36]
\draw[step=1,color=gray] (-3.5,-3.5) grid (3.5,3.5);
\draw[very thick,red] (0, 0) circle (.4);
\filldraw[black] (0,-1) circle (.4);
\filldraw[black] (0,+1) circle (.4);
\filldraw[black] (-1,0) circle (.4);
\filldraw[black] (+1,0) circle (.4);
\filldraw[black] ( 0,-2) circle (.4);
\filldraw[black] ( 0,+2) circle (.4);
\filldraw[black] (-2, 0) circle (.4);
\filldraw[black] (+2, 0) circle (.4);
\filldraw[black] (-2,-2) circle (.4);
\filldraw[black] (-2,+2) circle (.4);
\filldraw[black] (+2,-2) circle (.4);
\filldraw[black] (+2,+2) circle (.4);
\filldraw[black] (0,-3) circle (.4);
\filldraw[black] (0,+3) circle (.4);
\filldraw[black] (-3,0) circle (.4);
\filldraw[black] (+3,0) circle (.4);
\end{tikzpicture}
\end{subfigure}
\hfill
\begin{subfigure}[b]{0.14\textwidth}
\caption{\label{fig:1456nn}}
\begin{tikzpicture}[scale=0.36]
\draw[step=1,color=gray] (-3.5,-3.5) grid (3.5,3.5);
\draw[very thick,red] (0, 0) circle (.4);
\filldraw[black] (0,-1) circle (.4);
\filldraw[black] (0,+1) circle (.4);
\filldraw[black] (-1,0) circle (.4);
\filldraw[black] (+1,0) circle (.4);
\filldraw[black] (-2,-1) circle (.4);
\filldraw[black] (-2,+1) circle (.4);
\filldraw[black] (-1,-2) circle (.4);
\filldraw[black] (-1,+2) circle (.4);
\filldraw[black] (+1,-2) circle (.4);
\filldraw[black] (+1,+2) circle (.4);
\filldraw[black] (+2,-1) circle (.4);
\filldraw[black] (+2,+1) circle (.4);
\filldraw[black] (-2,-2) circle (.4);
\filldraw[black] (-2,+2) circle (.4);
\filldraw[black] (+2,-2) circle (.4);
\filldraw[black] (+2,+2) circle (.4);
\filldraw[black] (0,-3) circle (.4);
\filldraw[black] (0,+3) circle (.4);
\filldraw[black] (-3,0) circle (.4);
\filldraw[black] (+3,0) circle (.4);
\end{tikzpicture}
\end{subfigure}
\hfill
\begin{subfigure}[b]{0.14\textwidth}
\caption{\label{fig:2346nn}}
\begin{tikzpicture}[scale=0.36]
\draw[step=1,color=gray] (-3.5,-3.5) grid (3.5,3.5);
\draw[very thick,red] (0, 0) circle (.4);
\filldraw[black] (+1,-1) circle (.4);
\filldraw[black] (+1,+1) circle (.4);
\filldraw[black] (-1,-1) circle (.4);
\filldraw[black] (-1,+1) circle (.4);
\filldraw[black] ( 0,-2) circle (.4);
\filldraw[black] ( 0,+2) circle (.4);
\filldraw[black] (-2, 0) circle (.4);
\filldraw[black] (+2, 0) circle (.4);
\filldraw[black] (-2,-1) circle (.4);
\filldraw[black] (-2,+1) circle (.4);
\filldraw[black] (-1,-2) circle (.4);
\filldraw[black] (-1,+2) circle (.4);
\filldraw[black] (+1,-2) circle (.4);
\filldraw[black] (+1,+2) circle (.4);
\filldraw[black] (+2,-1) circle (.4);
\filldraw[black] (+2,+1) circle (.4);
\filldraw[black] (0,-3) circle (.4);
\filldraw[black] (0,+3) circle (.4);
\filldraw[black] (-3,0) circle (.4);
\filldraw[black] (+3,0) circle (.4);
\end{tikzpicture}
\end{subfigure}
\hfill
\begin{subfigure}[b]{0.14\textwidth}
\caption{\label{fig:2356nn}}
\begin{tikzpicture}[scale=0.36]
\draw[step=1,color=gray] (-3.5,-3.5) grid (3.5,3.5);
\draw[very thick,red] (0, 0) circle (.4);
\filldraw[black] (+1,-1) circle (.4);
\filldraw[black] (+1,+1) circle (.4);
\filldraw[black] (-1,-1) circle (.4);
\filldraw[black] (-1,+1) circle (.4);
\filldraw[black] ( 0,-2) circle (.4);
\filldraw[black] ( 0,+2) circle (.4);
\filldraw[black] (-2, 0) circle (.4);
\filldraw[black] (+2, 0) circle (.4);
\filldraw[black] (-2,-2) circle (.4);
\filldraw[black] (-2,+2) circle (.4);
\filldraw[black] (+2,-2) circle (.4);
\filldraw[black] (+2,+2) circle (.4);
\filldraw[black] (0,-3) circle (.4);
\filldraw[black] (0,+3) circle (.4);
\filldraw[black] (-3,0) circle (.4);
\filldraw[black] (+3,0) circle (.4);
\end{tikzpicture}
\end{subfigure}
\hfill
\begin{subfigure}[b]{0.14\textwidth}
\caption{\label{fig:2456nn}}
\begin{tikzpicture}[scale=0.36]
\draw[step=1,color=gray] (-3.5,-3.5) grid (3.5,3.5);
\draw[very thick,red] (0, 0) circle (.4);
\filldraw[black] (+1,-1) circle (.4);
\filldraw[black] (+1,+1) circle (.4);
\filldraw[black] (-1,-1) circle (.4);
\filldraw[black] (-1,+1) circle (.4);
\filldraw[black] (-2,-1) circle (.4);
\filldraw[black] (-2,+1) circle (.4);
\filldraw[black] (-1,-2) circle (.4);
\filldraw[black] (-1,+2) circle (.4);
\filldraw[black] (+1,-2) circle (.4);
\filldraw[black] (+1,+2) circle (.4);
\filldraw[black] (+2,-1) circle (.4);
\filldraw[black] (+2,+1) circle (.4);
\filldraw[black] (-2,-2) circle (.4);
\filldraw[black] (-2,+2) circle (.4);
\filldraw[black] (+2,-2) circle (.4);
\filldraw[black] (+2,+2) circle (.4);
\filldraw[black] (0,-3) circle (.4);
\filldraw[black] (0,+3) circle (.4);
\filldraw[black] (-3,0) circle (.4);
\filldraw[black] (+3,0) circle (.4);
\end{tikzpicture}
\end{subfigure}
\hfill
\begin{subfigure}[b]{0.14\textwidth}
\caption{\label{fig:3456nn}}
\begin{tikzpicture}[scale=0.36]
\draw[step=1,color=gray] (-3.5,-3.5) grid (3.5,3.5);
\draw[very thick,red] (0, 0) circle (.4);
\filldraw[black] ( 0,-2) circle (.4);
\filldraw[black] ( 0,+2) circle (.4);
\filldraw[black] (-2, 0) circle (.4);
\filldraw[black] (+2, 0) circle (.4);
\filldraw[black] (-2,-1) circle (.4);
\filldraw[black] (-2,+1) circle (.4);
\filldraw[black] (-1,-2) circle (.4);
\filldraw[black] (-1,+2) circle (.4);
\filldraw[black] (+1,-2) circle (.4);
\filldraw[black] (+1,+2) circle (.4);
\filldraw[black] (+2,-1) circle (.4);
\filldraw[black] (+2,+1) circle (.4);
\filldraw[black] (-2,-2) circle (.4);
\filldraw[black] (-2,+2) circle (.4);
\filldraw[black] (+2,-2) circle (.4);
\filldraw[black] (+2,+2) circle (.4);
\filldraw[black] (0,-3) circle (.4);
\filldraw[black] (0,+3) circle (.4);
\filldraw[black] (-3,0) circle (.4);
\filldraw[black] (+3,0) circle (.4);
\end{tikzpicture}
\end{subfigure}
\hfill
\begin{subfigure}[b]{0.14\textwidth}
\caption{\label{fig:12346nn}}
\begin{tikzpicture}[scale=0.36]
\draw[step=1,color=gray] (-3.5,-3.5) grid (3.5,3.5);
\draw[very thick,red] (0, 0) circle (.4);
\filldraw[black] (0,-1) circle (.4);
\filldraw[black] (0,+1) circle (.4);
\filldraw[black] (-1,0) circle (.4);
\filldraw[black] (+1,0) circle (.4);
\filldraw[black] (+1,-1) circle (.4);
\filldraw[black] (+1,+1) circle (.4);
\filldraw[black] (-1,-1) circle (.4);
\filldraw[black] (-1,+1) circle (.4);
\filldraw[black] ( 0,-2) circle (.4);
\filldraw[black] ( 0,+2) circle (.4);
\filldraw[black] (-2, 0) circle (.4);
\filldraw[black] (+2, 0) circle (.4);
\filldraw[black] (-2,-1) circle (.4);
\filldraw[black] (-2,+1) circle (.4);
\filldraw[black] (-1,-2) circle (.4);
\filldraw[black] (-1,+2) circle (.4);
\filldraw[black] (+1,-2) circle (.4);
\filldraw[black] (+1,+2) circle (.4);
\filldraw[black] (+2,-1) circle (.4);
\filldraw[black] (+2,+1) circle (.4);
\filldraw[black] (0,-3) circle (.4);
\filldraw[black] (0,+3) circle (.4);
\filldraw[black] (-3,0) circle (.4);
\filldraw[black] (+3,0) circle (.4);
\end{tikzpicture}
\end{subfigure}
\hfill
\begin{subfigure}[b]{0.14\textwidth}
\caption{\label{fig:12356nn}}
\begin{tikzpicture}[scale=0.36]
\draw[step=1,color=gray] (-3.5,-3.5) grid (3.5,3.5);
\draw[very thick,red] (0, 0) circle (.4);
\filldraw[black] (0,-1) circle (.4);
\filldraw[black] (0,+1) circle (.4);
\filldraw[black] (-1,0) circle (.4);
\filldraw[black] (+1,0) circle (.4);
\filldraw[black] (+1,-1) circle (.4);
\filldraw[black] (+1,+1) circle (.4);
\filldraw[black] (-1,-1) circle (.4);
\filldraw[black] (-1,+1) circle (.4);
\filldraw[black] ( 0,-2) circle (.4);
\filldraw[black] ( 0,+2) circle (.4);
\filldraw[black] (-2, 0) circle (.4);
\filldraw[black] (+2, 0) circle (.4);
\filldraw[black] (-2,-2) circle (.4);
\filldraw[black] (-2,+2) circle (.4);
\filldraw[black] (+2,-2) circle (.4);
\filldraw[black] (+2,+2) circle (.4);
\filldraw[black] (0,-3) circle (.4);
\filldraw[black] (0,+3) circle (.4);
\filldraw[black] (-3,0) circle (.4);
\filldraw[black] (+3,0) circle (.4);
\end{tikzpicture}
\end{subfigure}
\hfill
\begin{subfigure}[b]{0.14\textwidth}
\caption{\label{fig:12456nn}}
\begin{tikzpicture}[scale=0.36]
\draw[step=1,color=gray] (-3.5,-3.5) grid (3.5,3.5);
\draw[very thick,red] (0, 0) circle (.4);
\filldraw[black] (0,-1) circle (.4);
\filldraw[black] (0,+1) circle (.4);
\filldraw[black] (-1,0) circle (.4);
\filldraw[black] (+1,0) circle (.4);
\filldraw[black] (+1,-1) circle (.4);
\filldraw[black] (+1,+1) circle (.4);
\filldraw[black] (-1,-1) circle (.4);
\filldraw[black] (-1,+1) circle (.4);
\filldraw[black] (-2,-1) circle (.4);
\filldraw[black] (-2,+1) circle (.4);
\filldraw[black] (-1,-2) circle (.4);
\filldraw[black] (-1,+2) circle (.4);
\filldraw[black] (+1,-2) circle (.4);
\filldraw[black] (+1,+2) circle (.4);
\filldraw[black] (+2,-1) circle (.4);
\filldraw[black] (+2,+1) circle (.4);
\filldraw[black] (-2,-2) circle (.4);
\filldraw[black] (-2,+2) circle (.4);
\filldraw[black] (+2,-2) circle (.4);
\filldraw[black] (+2,+2) circle (.4);
\filldraw[black] (0,-3) circle (.4);
\filldraw[black] (0,+3) circle (.4);
\filldraw[black] (-3,0) circle (.4);
\filldraw[black] (+3,0) circle (.4);
\end{tikzpicture}
\end{subfigure}
\hfill
\begin{subfigure}[b]{0.14\textwidth}
\caption{\label{fig:13456nn}}
\begin{tikzpicture}[scale=0.36]
\draw[step=1,color=gray] (-3.5,-3.5) grid (3.5,3.5);
\draw[very thick,red] (0, 0) circle (.4);
\filldraw[black] (0,-1) circle (.4);
\filldraw[black] (0,+1) circle (.4);
\filldraw[black] (-1,0) circle (.4);
\filldraw[black] (+1,0) circle (.4);
\filldraw[black] ( 0,-2) circle (.4);
\filldraw[black] ( 0,+2) circle (.4);
\filldraw[black] (-2, 0) circle (.4);
\filldraw[black] (+2, 0) circle (.4);
\filldraw[black] (-2,-1) circle (.4);
\filldraw[black] (-2,+1) circle (.4);
\filldraw[black] (-1,-2) circle (.4);
\filldraw[black] (-1,+2) circle (.4);
\filldraw[black] (+1,-2) circle (.4);
\filldraw[black] (+1,+2) circle (.4);
\filldraw[black] (+2,-1) circle (.4);
\filldraw[black] (+2,+1) circle (.4);
\filldraw[black] (-2,-2) circle (.4);
\filldraw[black] (-2,+2) circle (.4);
\filldraw[black] (+2,-2) circle (.4);
\filldraw[black] (+2,+2) circle (.4);
\filldraw[black] (0,-3) circle (.4);
\filldraw[black] (0,+3) circle (.4);
\filldraw[black] (-3,0) circle (.4);
\filldraw[black] (+3,0) circle (.4);
\end{tikzpicture}
\end{subfigure}
\hfill
\begin{subfigure}[b]{0.14\textwidth}
\caption{\label{fig:23456nn}}
\begin{tikzpicture}[scale=0.36]
\draw[step=1,color=gray] (-3.5,-3.5) grid (3.5,3.5);
\draw[very thick,red] (0, 0) circle (.4);
\filldraw[black] (+1,-1) circle (.4);
\filldraw[black] (+1,+1) circle (.4);
\filldraw[black] (-1,-1) circle (.4);
\filldraw[black] (-1,+1) circle (.4);
\filldraw[black] ( 0,-2) circle (.4);
\filldraw[black] ( 0,+2) circle (.4);
\filldraw[black] (-2, 0) circle (.4);
\filldraw[black] (+2, 0) circle (.4);
\filldraw[black] (-2,-1) circle (.4);
\filldraw[black] (-2,+1) circle (.4);
\filldraw[black] (-1,-2) circle (.4);
\filldraw[black] (-1,+2) circle (.4);
\filldraw[black] (+1,-2) circle (.4);
\filldraw[black] (+1,+2) circle (.4);
\filldraw[black] (+2,-1) circle (.4);
\filldraw[black] (+2,+1) circle (.4);
\filldraw[black] (-2,-2) circle (.4);
\filldraw[black] (-2,+2) circle (.4);
\filldraw[black] (+2,-2) circle (.4);
\filldraw[black] (+2,+2) circle (.4);
\filldraw[black] (0,-3) circle (.4);
\filldraw[black] (0,+3) circle (.4);
\filldraw[black] (-3,0) circle (.4);
\filldraw[black] (+3,0) circle (.4);
\end{tikzpicture}
\end{subfigure}
\hfill
\begin{subfigure}[b]{0.14\textwidth}
\caption{\label{fig:123456nn}}
\begin{tikzpicture}[scale=0.36]
\draw[step=1,color=gray] (-3.5,-3.5) grid (3.5,3.5);
\draw[very thick,red] (0, 0) circle (.4);
\filldraw[black] (0,-1) circle (.4);
\filldraw[black] (0,+1) circle (.4);
\filldraw[black] (-1,0) circle (.4);
\filldraw[black] (+1,0) circle (.4);
\filldraw[black] (+1,-1) circle (.4);
\filldraw[black] (+1,+1) circle (.4);
\filldraw[black] (-1,-1) circle (.4);
\filldraw[black] (-1,+1) circle (.4);
\filldraw[black] ( 0,-2) circle (.4);
\filldraw[black] ( 0,+2) circle (.4);
\filldraw[black] (-2, 0) circle (.4);
\filldraw[black] (+2, 0) circle (.4);
\filldraw[black] (-2,-1) circle (.4);
\filldraw[black] (-2,+1) circle (.4);
\filldraw[black] (-1,-2) circle (.4);
\filldraw[black] (-1,+2) circle (.4);
\filldraw[black] (+1,-2) circle (.4);
\filldraw[black] (+1,+2) circle (.4);
\filldraw[black] (+2,-1) circle (.4);
\filldraw[black] (+2,+1) circle (.4);
\filldraw[black] (-2,-2) circle (.4);
\filldraw[black] (-2,+2) circle (.4);
\filldraw[black] (+2,-2) circle (.4);
\filldraw[black] (+2,+2) circle (.4);
\filldraw[black] (0,-3) circle (.4);
\filldraw[black] (0,+3) circle (.4);
\filldraw[black] (-3,0) circle (.4);
\filldraw[black] (+3,0) circle (.4);
\end{tikzpicture}
\end{subfigure}
\caption{\label{fig:sq-neighbourhoods}Shapes of neighborhoods on square lattice combined with basics neighborhoods presented in \Cref{fig:sq-basic-neighbourhoods} and containing sites from the sixth coordination zone (\Cref{fig:sq-6nn}).
\subref{fig:16nn}     \textsc{sq-1,6},
\subref{fig:26nn}     \textsc{sq-2,6},
\subref{fig:36nn}     \textsc{sq-3,6},
\subref{fig:46nn}     \textsc{sq-4,6},
\subref{fig:56nn}     \textsc{sq-5,6},
\subref{fig:126nn}    \textsc{sq-1,2,6},
\subref{fig:136nn}    \textsc{sq-1,3,6},
\subref{fig:146nn}    \textsc{sq-1,4,6},
\subref{fig:156nn}    \textsc{sq-1,5,6},
\subref{fig:236nn}    \textsc{sq-2,3,6},
\subref{fig:246nn}    \textsc{sq-2,4,6},
\subref{fig:256nn}    \textsc{sq-2,5,6},
\subref{fig:346nn}    \textsc{sq-3,4,6},
\subref{fig:356nn}    \textsc{sq-3,5,6},
\subref{fig:456nn}    \textsc{sq-4,5,6},
\subref{fig:1236nn}   \textsc{sq-1,2,3,6},
\subref{fig:1246nn}   \textsc{sq-1,2,4,6},
\subref{fig:1256nn}   \textsc{sq-1,2,5,6},
\subref{fig:1346nn}   \textsc{sq-1,3,4,6},
\subref{fig:1356nn}   \textsc{sq-1,3,5,6},
\subref{fig:1456nn}   \textsc{sq-1,4,5,6},
\subref{fig:2346nn}   \textsc{sq-2,3,4,6},
\subref{fig:2356nn}   \textsc{sq-2,3,5,6},
\subref{fig:2456nn}   \textsc{sq-2,4,5,6},
\subref{fig:3456nn}   \textsc{sq-3,4,5,6},
\subref{fig:12346nn}  \textsc{sq-1,2,3,4,6},
\subref{fig:12356nn}  \textsc{sq-1,2,3,5,6},
\subref{fig:12456nn}  \textsc{sq-1,2,4,5,6},
\subref{fig:13456nn}  \textsc{sq-1,3,4,5,6},
\subref{fig:23456nn}  \textsc{sq-2,3,4,5,6},
\subref{fig:123456nn} \textsc{sq-1,2,3,4,5,6}}
\end{figure*}

\newpage
\section*{Supplemental Material}
\appendix

\section{\label{app:functions}Boundaries procedures}

Below, we present a set of {\tt boundaries()} functions (written in C) to be replaced in the Newman--Ziff program published in Reference \onlinecite{NewmanZiff2001} to obtain the single realization of $\mathcal S_{\max}(n;L)$ for the neighborhoods presented in \Cref{fig:sq-1nn,fig:sq-2nn,fig:sq-3nn,fig:sq-4nn,fig:sq-5nn,fig:sq-6nn}.

\subsection{{SQ-1}}
{\tt boundaries()} function for the \textsc{sq-1} neighborhood (originally presented in Reference \onlinecite{NewmanZiff2001}).

\begin{lstlisting}[language=C]
void boundaries()
{
  int i,j;
  for (i=0; i<N; i++) {
// 1nn core:
    nn[i][0] = (N+i +1)%N;
    nn[i][1] = (N+i -1)%N;
    nn[i][2] = (N+i +L)%N;
    nn[i][3] = (N+i -L)%N;
// 1nn left border:
    if (i%L==0)     nn[i][1] = (N+i+L -1)%N;
// 1nn right border:
    if ((i+1)%L==0) nn[i][0] = (N+i-L +1)%N;
  }
}
\end{lstlisting}

\subsection{{SQ-2}}
{\tt boundaries()} function for the \textsc{sq-2} neighborhood.

\begin{lstlisting}[language=C]
void boundaries()
{
  int i,j;
  for (i=0; i<N; i++) {
// 2nn core:
    nn[i][0] = (N+i +L+1)%N;
    nn[i][1] = (N+i +L-1)%N;
    nn[i][2] = (N+i -L+1)%N;
    nn[i][3] = (N+i -L-1)%N;
// 2nn left border:
    if(i%L==0) {
       nn[i][1] = (N+i+L +L-1)%N;
       nn[i][3] = (N+i+L -L-1)%N; }
// 2nn right border:
    if((i+1)%L==0) {
       nn[i][0] = (N+i-L +L+1)%N;
       nn[i][2] = (N+i-L -L+1)%N; }
  }
}
\end{lstlisting}

\subsection{{SQ-3}}
{\tt boundaries()} function for the \textsc{sq-3} neighborhood.

\begin{lstlisting}[language=C]
void boundaries()
{
  int i,j;
  for (i=0; i<N; i++) {
// 3nn core:
    nn[i][0] = (N+i +2*L)%N;
    nn[i][1] = (N+i -2*L)%N;
    nn[i][2] = (N+i +2)%N;
    nn[i][3] = (N+i -2)%N;
// 3nn left border:
    if(i%L==0 || i%L==1) 
      nn[i][3] = (N+i+L -2)%N;
// 3nn right border:
    if((i+1)%L==0 || (i+2)%L==0)
      nn[i][2] = (N+i-L +2)%N;
  }
}
\end{lstlisting}

\subsection{{SQ-4}}
{\tt boundaries()} function for the neighborhood \textsc{sq-4}.
The preprocesor directive {\tt \#define Z 4} in source code in Reference \onlinecite{NewmanZiff2001} requires replacing to {\tt \#define Z 8}.

\begin{lstlisting}[language=C]
void boundaries()
{
  int i,j;
  for (i=0; i<N; i++) {
   // 4nn core:
      nn[i][0] = (N+i +L+2)%N;
      nn[i][1] = (N+i +L-2)%N;
      nn[i][2] = (N+i -L+2)%N;
      nn[i][3] = (N+i -L-2)%N;
      nn[i][4] = (N+i +2*L+1)%N;
      nn[i][5] = (N+i +2*L-1)%N;
      nn[i][6] = (N+i -2*L+1)%N;
      nn[i][7] = (N+i -2*L-1)%N;
   // 4nn left border:
      if(i%L==0) {
        nn[i][1] = (N+i+L +L-2)%N;
        nn[i][3] = (N+i+L -L-2)%N;
        nn[i][5] = (N+i+L +2*L-1)%N;
        nn[i][7] = (N+i+L -2*L-1)%N; }
      if(i%L==1) {
        nn[i][1] = (N+i+L +L-2)%N;
        nn[i][3] = (N+i+L -L-2)%N; }
   // 4nn right border:
      if((i+1)%L==0) {
        nn[i][0] = (N+i-L +L+2)%N;
        nn[i][2] = (N+i-L -L+2)%N;
        nn[i][4] = (N+i-L +2*L+1)%N;
        nn[i][6] = (N+i-L -2*L+1)%N; }
      if((i+2)%L==0) {
        nn[i][0] = (N+i-L +L+2)%N;
        nn[i][2] = (N+i-L -L+2)%N; }
  }
}
\end{lstlisting}

\subsection{{SQ-5}}
{\tt boundaries()} function for the \textsc{sq-5} neighborhood.

\begin{lstlisting}[language=C]
void boundaries()
{
  int i,j;
  for (i=0; i<N; i++) {
// 5nn core:
    nn[i][0] = (N+i +2*L+2)%N;
    nn[i][1] = (N+i +2*L-2)%N;
    nn[i][2] = (N+i -2*L+2)%N;
    nn[i][3] = (N+i -2*L-2)%N;
// 5nn left border:
    if(i%L==0 || i%L==1) {
       nn[i][1] = (N+i+L +2*L-2)%N;
       nn[i][3] = (N+i+L -2*L-2)%N; }
// 5nn right border:
    if((i+1)%L==0 || (i+2)%L==0) {
       nn[i][0] = (N+i-L +2*L+2)%N;
       nn[i][2] = (N+i-L -2*L+2)%N; }
  }
}
\end{lstlisting}

\subsection{{SQ-6}}
{\tt boundaries()} function for the \textsc{sq-6} neighborhood.

\begin{lstlisting}[language=C]
void boundaries()
{
  int i,j;
  for (i=0; i<N; i++) {
// 6nn core:
    nn[i][0] = (N+i +3*L)%N;
    nn[i][1] = (N+i -3*L)%N;
    nn[i][2] = (N+i +3)%N;
    nn[i][3] = (N+i -3)%N;
// 6nn left border:
    if(i%L==0 || i%L==1 || i%L==2)
      nn[i][3] = (N+i+L -3)%N;
// 6nn right border:
    if((i+1)%L==0 || (i+2)%L==0 || (i+3)%L==0)
      nn[i][2] = (N+i-L +3)%N;
  }
}
\end{lstlisting}

\section{\label{app:resuts}Dependencies of $\mathcal{P}_{\max}\cdot L^{\beta/\nu}$ on the probability of occupation $p$}

\Cref{fig:PmaxLbetanuvsp} presents the dependencies of $\mathcal{P}_{\max}\cdot L^{\beta/\nu}$ on the probability of occupation $p$ for neighborhoods ranging from \textsc{sq-6} to \textsc{sq-1,2,3,4,5,6} for various linear system sizes $L=128$, 256, 512, 1024, 2048, and 4096.

\begin{figure*}[htbp]
\begin{subfigure}[b]{0.32\textwidth}
\caption{\label{fig:PmaxLbetanuvsp-SQ-6}}
\includegraphics[width=0.99\textwidth]{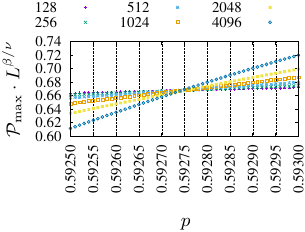}
\end{subfigure}
\hfill 
\begin{subfigure}[b]{0.32\textwidth}
\caption{\label{fig:PmaxLbetanuvsp-SQ-16}}
\includegraphics[width=0.99\textwidth]{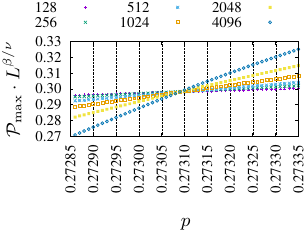}
\end{subfigure}
\hfill 
\begin{subfigure}[b]{0.32\textwidth}
\caption{\label{fig:PmaxLbetanuvsp-SQ-26}}
\includegraphics[width=0.99\textwidth]{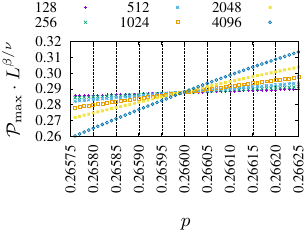}
\end{subfigure}
\hfill 
\begin{subfigure}[b]{0.32\textwidth}
\caption{\label{fig:PmaxLbetanuvsp-SQ-36}}
\includegraphics[width=0.99\textwidth]{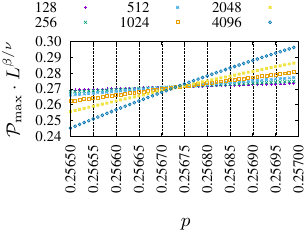}
\end{subfigure}
\hfill 
\begin{subfigure}[b]{0.32\textwidth}
\caption{\label{fig:PmaxLbetanuvsp-SQ-46}}
\includegraphics[width=0.99\textwidth]{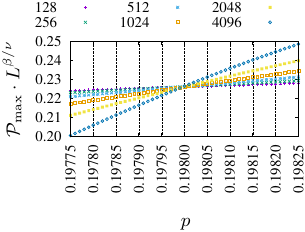}
\end{subfigure}
\hfill 
\begin{subfigure}[b]{0.32\textwidth}
\caption{\label{fig:PmaxLbetanuvsp-SQ-56}}
\includegraphics[width=0.99\textwidth]{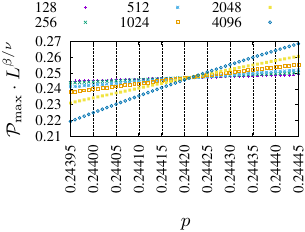}
\end{subfigure}
\hfill 
\begin{subfigure}[b]{0.32\textwidth}
\caption{\label{fig:PmaxLbetanuvsp-SQ-126}}
\includegraphics[width=0.99\textwidth]{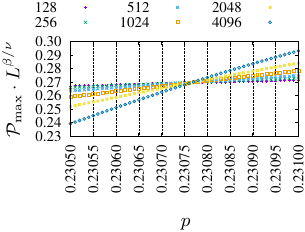}
\end{subfigure}
\hfill 
\begin{subfigure}[b]{0.32\textwidth}
\caption{\label{fig:PmaxLbetanuvsp-SQ-136}}
\includegraphics[width=0.99\textwidth]{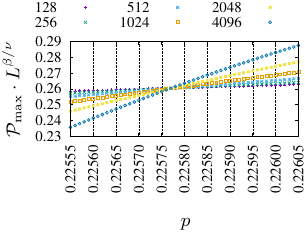}
\end{subfigure}
\begin{subfigure}[b]{0.32\textwidth}
\caption{\label{fig:PmaxLbetanuvsp-SQ-146}}
\includegraphics[width=0.99\textwidth]{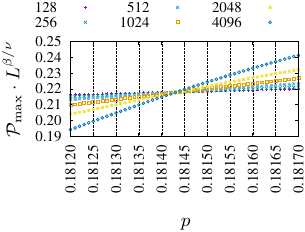}
\end{subfigure}
\hfill 
\begin{subfigure}[b]{0.32\textwidth}
\caption{\label{fig:PmaxLbetanuvsp-SQ-156}}
\includegraphics[width=0.99\textwidth]{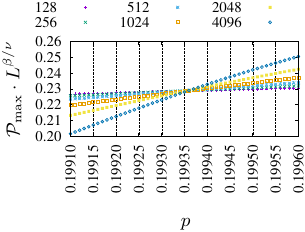}
\end{subfigure}
\hfill 
\begin{subfigure}[b]{0.32\textwidth}
\caption{\label{fig:PmaxLbetanuvsp-SQ-236}}
\includegraphics[width=0.99\textwidth]{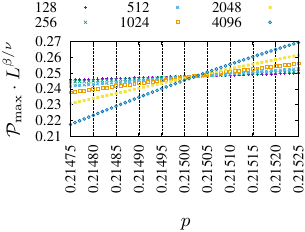}
\end{subfigure}
\hfill 
\begin{subfigure}[b]{0.32\textwidth}
\caption{\label{fig:PmaxLbetanuvsp-SQ-246}}
\includegraphics[width=0.99\textwidth]{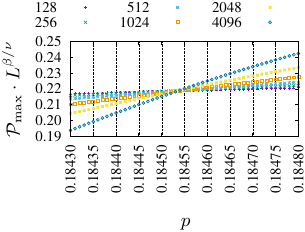}
\end{subfigure}
\hfill 
\begin{subfigure}[b]{0.32\textwidth}
\caption{\label{fig:PmaxLbetanuvsp-SQ-256}}
\includegraphics[width=0.99\textwidth]{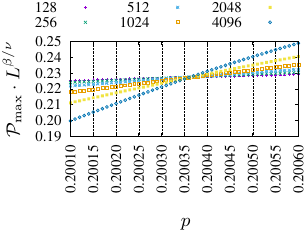}
\end{subfigure}
\hfill 
\begin{subfigure}[b]{0.32\textwidth}
\caption{\label{fig:PmaxLbetanuvsp-SQ-346}}
\includegraphics[width=0.99\textwidth]{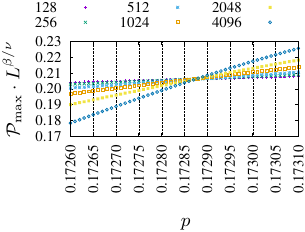}
\end{subfigure}
\hfill 
\begin{subfigure}[b]{0.32\textwidth}
\caption{\label{fig:PmaxLbetanuvsp-SQ-356}}
\includegraphics[width=0.99\textwidth]{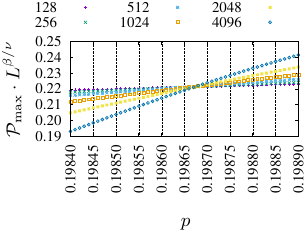}
\end{subfigure}
\end{figure*}
\begin{figure*}\ContinuedFloat
\begin{subfigure}[b]{0.32\textwidth}
\caption{\label{fig:PmaxLbetanuvsp-SQ-456}}
\includegraphics[width=0.99\textwidth]{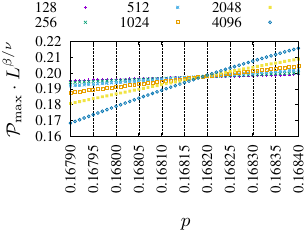}
\end{subfigure}
\hfill 
\begin{subfigure}[b]{0.32\textwidth}
\caption{\label{fig:PmaxLbetanuvsp-SQ-1236}}
\includegraphics[width=0.99\textwidth]{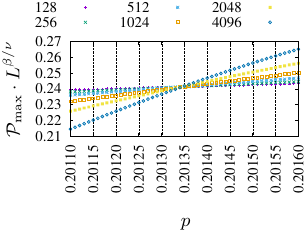}
\end{subfigure}
\hfill 
\begin{subfigure}[b]{0.32\textwidth}
\caption{\label{fig:PmaxLbetanuvsp-SQ-1246}}
\includegraphics[width=0.99\textwidth]{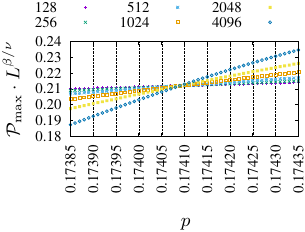}
\end{subfigure}
\hfill 
\begin{subfigure}[b]{0.32\textwidth}
\caption{\label{fig:PmaxLbetanuvsp-SQ-1256}}
\includegraphics[width=0.99\textwidth]{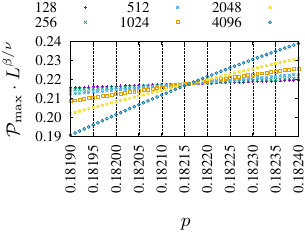}
\end{subfigure}
\hfill 
\begin{subfigure}[b]{0.32\textwidth}
\caption{\label{fig:PmaxLbetanuvsp-SQ-1346}}
\includegraphics[width=0.99\textwidth]{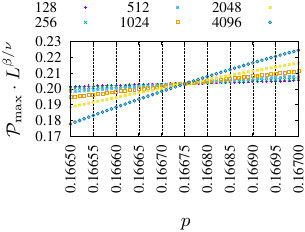}
\end{subfigure}
\hfill 
\begin{subfigure}[b]{0.32\textwidth}
\caption{\label{fig:PmaxLbetanuvsp-SQ-1356}}
\includegraphics[width=0.99\textwidth]{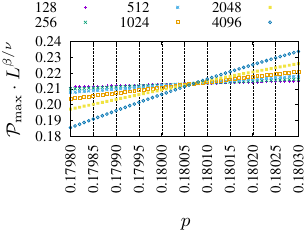}
\end{subfigure}
\hfill 
\begin{subfigure}[b]{0.32\textwidth}
\caption{\label{fig:PmaxLbetanuvsp-SQ-1456}}
\includegraphics[width=0.99\textwidth]{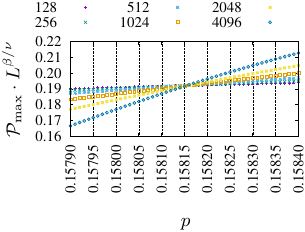}
\end{subfigure}
\hfill 
\begin{subfigure}[b]{0.32\textwidth}
\caption{\label{fig:PmaxLbetanuvsp-SQ-2346}}
\includegraphics[width=0.99\textwidth]{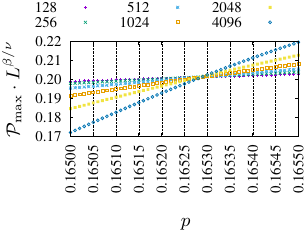}
\end{subfigure}
\hfill 
\begin{subfigure}[b]{0.32\textwidth}
\caption{\label{fig:PmaxLbetanuvsp-SQ-2356}}
\includegraphics[width=0.99\textwidth]{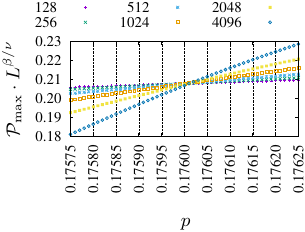}
\end{subfigure}
\hfill 
\begin{subfigure}[b]{0.32\textwidth}
\caption{\label{fig:PmaxLbetanuvsp-SQ-2456}}
\includegraphics[width=0.99\textwidth]{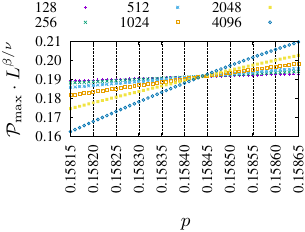}
\end{subfigure}
\hfill 
\begin{subfigure}[b]{0.32\textwidth}
\caption{\label{fig:PmaxLbetanuvsp-SQ-3456}}
\includegraphics[width=0.99\textwidth]{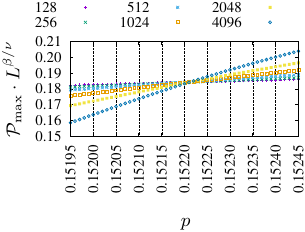}
\end{subfigure}
\hfill 
\begin{subfigure}[b]{0.32\textwidth}
\caption{\label{fig:PmaxLbetanuvsp-SQ-12346}}
\includegraphics[width=0.99\textwidth]{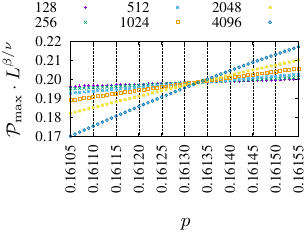}
\end{subfigure}
\hfill 
\begin{subfigure}[b]{0.32\textwidth}
\caption{\label{fig:PmaxLbetanuvsp-SQ-12356}}
\includegraphics[width=0.99\textwidth]{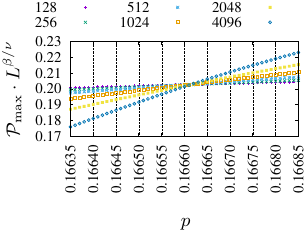}
\end{subfigure}
\hfill 
\begin{subfigure}[b]{0.32\textwidth}
\caption{\label{fig:PmaxLbetanuvsp-SQ-12456}}
\includegraphics[width=0.99\textwidth]{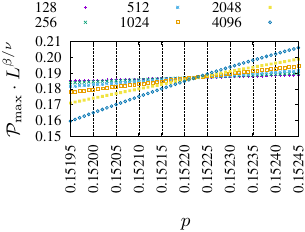}
\end{subfigure}
\hfill 
\begin{subfigure}[b]{0.32\textwidth}
\caption{\label{fig:PmaxLbetanuvsp-SQ-13456}}
\includegraphics[width=0.99\textwidth]{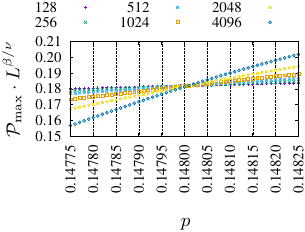}
\end{subfigure}
\end{figure*}
\begin{figure*}\ContinuedFloat
\begin{subfigure}[b]{0.32\textwidth}
\caption{\label{fig:PmaxLbetanuvsp-SQ-23456}}
\includegraphics[width=0.99\textwidth]{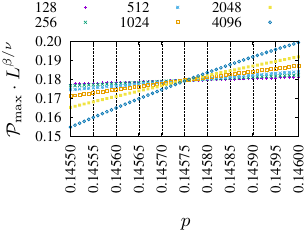}
\end{subfigure}
\hfill 
\begin{subfigure}[b]{0.32\textwidth}
\caption{\label{fig:PmaxLbetanuvsp-SQ-123456}}
\includegraphics[width=0.99\textwidth]{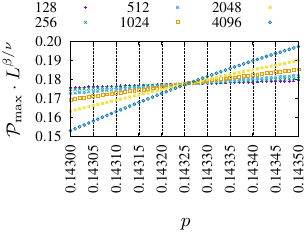}
\end{subfigure}
\caption{\label{fig:PmaxLbetanuvsp}$L^{\varepsilon_1}\mathcal P_\text{max}$ vs. $p$ for various complex neighborhoods and various values of $L$ indicated in the title of the figures. The results are averaged over the realizations of the $R=10^5$ system. The probability of occupation $p$ is scanned with $\Delta p=10^{-5}$ separation step.
\subref{fig:PmaxLbetanuvsp-SQ-6}      \textsc{sq-6},
\subref{fig:PmaxLbetanuvsp-SQ-16}     \textsc{sq-1,6},
\subref{fig:PmaxLbetanuvsp-SQ-26}     \textsc{sq-2,6},
\subref{fig:PmaxLbetanuvsp-SQ-36}     \textsc{sq-3,6},
\subref{fig:PmaxLbetanuvsp-SQ-46}     \textsc{sq-4,6},
\subref{fig:PmaxLbetanuvsp-SQ-56}     \textsc{sq-5,6},
\subref{fig:PmaxLbetanuvsp-SQ-126}    \textsc{sq-1,2,6},
\subref{fig:PmaxLbetanuvsp-SQ-136}    \textsc{sq-1,3,6},
\subref{fig:PmaxLbetanuvsp-SQ-146}    \textsc{sq-1,4,6},
\subref{fig:PmaxLbetanuvsp-SQ-156}    \textsc{sq-1,5,6},
\subref{fig:PmaxLbetanuvsp-SQ-236}    \textsc{sq-2,3,6},
\subref{fig:PmaxLbetanuvsp-SQ-246}    \textsc{sq-2,4,6},
\subref{fig:PmaxLbetanuvsp-SQ-256}    \textsc{sq-2,5,6},
\subref{fig:PmaxLbetanuvsp-SQ-346}    \textsc{sq-3,4,6},
\subref{fig:PmaxLbetanuvsp-SQ-356}    \textsc{sq-3,5,6},
\subref{fig:PmaxLbetanuvsp-SQ-456}    \textsc{sq-4,5,6},
\subref{fig:PmaxLbetanuvsp-SQ-1236}   \textsc{sq-1,2,3,6},
\subref{fig:PmaxLbetanuvsp-SQ-1246}   \textsc{sq-1,2,4,6},
\subref{fig:PmaxLbetanuvsp-SQ-1256}   \textsc{sq-1,2,5,6},
\subref{fig:PmaxLbetanuvsp-SQ-1346}   \textsc{sq-1,3,4,6},
\subref{fig:PmaxLbetanuvsp-SQ-1356}   \textsc{sq-1,3,5,6},
\subref{fig:PmaxLbetanuvsp-SQ-1456}   \textsc{sq-1,4,5,6},
\subref{fig:PmaxLbetanuvsp-SQ-2346}   \textsc{sq-2,3,4,6},
\subref{fig:PmaxLbetanuvsp-SQ-2356}   \textsc{sq-2,3,5,6},
\subref{fig:PmaxLbetanuvsp-SQ-2456}   \textsc{sq-2,4,5,6},
\subref{fig:PmaxLbetanuvsp-SQ-3456}   \textsc{sq-3,4,5,6},
\subref{fig:PmaxLbetanuvsp-SQ-12346}  \textsc{sq-1,2,3,4,6},
\subref{fig:PmaxLbetanuvsp-SQ-12356}  \textsc{sq-1,2,3,5,6},
\subref{fig:PmaxLbetanuvsp-SQ-12456}  \textsc{sq-1,2,4,5,6},
\subref{fig:PmaxLbetanuvsp-SQ-13456}  \textsc{sq-1,3,4,5,6},
\subref{fig:PmaxLbetanuvsp-SQ-23456}  \textsc{sq-2,3,4,5,6},
\subref{fig:PmaxLbetanuvsp-SQ-123456} \textsc{sq-1,2,3,4,5,6}}
\end{figure*}

\end{document}